\newif\ifSPACEHACK
\SPACEHACKfalse

\newif\ifDEBUG
\DEBUGfalse
\DEBUGtrue

\newif\ifANONYMOUS
\ANONYMOUSfalse

\newif\ifEXTENDED
 \EXTENDEDfalse

\newif\ifARXIV
\ARXIVfalse
\ARXIVtrue

\newif\ifSHOWREVISIONS
\SHOWREVISIONStrue

\ifANONYMOUS
  \documentclass[acmsmall,screen,review,anonymous]{template/acmart}
\else
  \documentclass[acmsmall,screen, conf]{template/acmart}
\fi

\setcopyright{none}

\AtBeginDocument{%
  }

\usepackage{amsmath,amsfonts}
\usepackage{algorithmic}
\usepackage{graphicx}
\usepackage{textcomp}
\usepackage{xcolor}
\usepackage{booktabs} 
\usepackage{xspace} 
\usepackage[normalem]{ulem}
\usepackage{makecell}
\usepackage{tcolorbox}
\usepackage{enumitem}
\usepackage{siunitx}
\usepackage[vskip=1em,font=itshape,leftmargin=2em,rightmargin=2em]{quoting}
\usepackage{svg}
\usepackage{listings}
\usepackage{subfigure}
\usepackage{xcolor}
\PassOptionsToPackage{hyphens}{url}\usepackage{hyperref}

\usepackage{soul}
\usepackage{hyperref}
\usepackage{cleveref}

\ifDEBUG
   \newcommand{\JD}[1]{\textcolor{purple}{[Jamie says: #1]}}
    \newcommand{\KC}[1]{\textcolor{blue}{[Kelechi says: #1]}}
    \newcommand{\SC}[1]{\textcolor{olive}{[Sophie says: #1]}}
    
    \newcommand{\ST}[1]{\textcolor{orange}{[Santiago says: #1]}}
    \newcommand{\SO}[1]{\textcolor{pink}{[Sofia says: #1]}}
    
    \newcommand{\SJ}[1]{\textcolor{red}{[Sooyeon says: #1]}}
    
\else
    \newcommand{\JD}[1]{}
    \newcommand{\KC}[1]{}
    \newcommand{\SC}[1]{}
    \newcommand{\AK}[1]{}
    \newcommand{\ST}[1]{}
    \newcommand{\SO}[1]{}
    \newcommand{\SJ}[1]{}
    
\fi

\newcommand{\myparagraph}[1]{\vspace{0.1cm}\textbf{#1}}

\usepackage{balance} 

\usepackage{amsmath}
\usepackage{cleveref}

\crefformat{section}{\S#2#1#3}
\crefname{figure}{Figure}{Figures}
\crefname{appendix}{Appendix}{Appendices}
\crefname{table}{Table}{Tables}
\crefname{table2}{Table}{Tables}
\crefname{algorithm}{Algorithm}{Algorithms}
\crefname{listing}{Listing}{Listings}
\crefname{theorem}{Theorem}{Theorems}
\crefname{thm}{Theorem}{Theorems}
\crefname{lemma}{Lemma}{Lemmata}
\crefname{equation}{Eq.}{Eqs.}
\crefformat{Grammar}{Grammar #1}

\usepackage{enumitem}
\usepackage{booktabs}

\usepackage{makecell}
\usepackage{multirow}
\ifdefined\XeTeXversion
\else
  \usepackage[T1]{fontenc}
\fi

\newcommand{\ie}{\textit{i.e.,} }
\newcommand{\eg}{\textit{e.g.,} }
\newcommand{\etal}{\textit{et al.}\xspace}

\newcommand{\issue}{issues and discussions\xspace}

\usepackage{enumitem}

\usepackage{quoting}

\quotingsetup{
  leftmargin=0.15cm,
  rightmargin=0.15cm,
  font=italic,
  vskip=0.10cm,        
  indentfirst=false
}

\ifdefined\DeclareUnicodeCharacter
  \DeclareUnicodeCharacter{2060}{\nolinebreak}
\fi

\newcounter{finding}

\ifSHOWREVISIONS
  \DeclareRobustCommand{\rev}[1]{{\color{ACMBlue}#1}}
\else
  \DeclareRobustCommand{\rev}[1]{#1}
\fi

\usepackage{tikz}
\usepackage{tabularx}
\usepackage{longtable}
\usepackage{filecontents}
\usepackage{placeins}

\tcbset{
  lessonbox/.style={
    colback=violet!6!white,
    colframe=violet!35!black,
    boxrule=0.5pt,
    arc=2pt,
    left=6pt,
    right=6pt,
    top=4pt,
    bottom=4pt,
    boxsep=2pt,
    before skip=9pt,
    after skip=9pt
  }
}

\tcbset{
  findingsbox/.style={
    colback=green!6!white,
    colframe=green!40!black,
    boxrule=0.5pt,
    arc=2pt,
    left=6pt,
    right=6pt,
    top=4pt,
    bottom=4pt,
    boxsep=2pt,
    before skip=7pt,
    after skip=7pt
  }
}

\begin{document}

\newcommand{\mytitle}{}
\renewcommand{\mytitle}{Signed, Struggled, Delivered: The Usability Life-Cycle of Identity-Based Software Signing}
\renewcommand{\mytitle}{A Longitudinal Study of Usability in Identity-Based Software Signing}
\renewcommand{\mytitle}{From Key Management to Coordination Burden: Usability Challenges in Identity-Based Software Signing Tools}
\renewcommand{\mytitle}{Identity-Based Signing Shifts Usability Burdens Rather Than Solving Them: Evidence from Five Signing Tools}
\renewcommand{\mytitle}{Identity-Based Signing Shifts, Not Solves, Usability: Evidence from Five Identity-Based Signing Ecosystems}

\title{\mytitle}

\ifANONYMOUS
  \author{Anonymous Author(s)}
  \affiliation{%
    \institution{Anonymous Institution}
    \country{Anonymous}}
\else
  \author{Kelechi G.\ Kalu}
  \email{kalu@purdue.edu}
  \affiliation{%
    \institution{Purdue University}
    \city{West Lafayette}
    \state{Indiana}
    \country{USA}}

  \author{Hieu Tran}
  \email{tran335@purdue.edu}
  \affiliation{%
    \institution{Purdue University}
    \city{West Lafayette}
    \state{Indiana}
    \country{USA}}

  \author{Santiago Torres-Arias}
  \email{santiagotorres@purdue.edu}
  \affiliation{%
    \institution{Purdue University}
    \city{West Lafayette}
    \state{Indiana}
    \country{USA}}

  \author{Sooyeon Jeong}
  \email{sooyeonj@purdue.edu}
  \affiliation{%
    \institution{Purdue University}
    \city{West Lafayette}
    \state{Indiana}
    \country{USA}}

  \author{James C.\ Davis}
  \email{davisjam@purdue.edu}
  \affiliation{%
    \institution{Purdue University}
    \city{West Lafayette}
    \state{Indiana}
    \country{USA}}
\fi

\begin{abstract}
Identity-based software signing tools aim to make software artifact provenance verifiable while reducing the operational burden of long-lived key management.
These tools are becoming critical infrastructure for secure software supply chains and are increasingly embedded in build, deployment, and release workflows.
However, there is limited cross-tool, longitudinal evidence about which usability problems arise in practice and how reported concerns change over time.
This gap matters because unusable signing and verification workflows can cause incomplete adoption, misconfiguration, or skipped verification, undermining intended integrity guarantees.

We conducted the first mining-software-repositories study of five identity-based signing ecosystems with public development repositories: Sigstore, OpenPubKey, HashiCorp Vault, Keyfactor, and Notary~v2.
We analyzed approximately 3{,}900 GitHub issue discussions from November~2021 through November~2025.
We coded each issue for the reported usability concern and affected surface, comparing patterns across tools and over time.
Our results show that across ecosystems, reported concerns are especially prominent around CLI tooling, policy/configuration, and verification workflows.
Using longitudinal Poisson trend analysis, we characterize how reported usability concerns change over time.
We found that reported issue counts decline substantially for several repositories, increase for OpenPubKey, and show no statistically detectable overall trend for the Keyfactor repositories, while trajectories also vary across concern types.
These results indicate that identity-based signing reduces direct key-management burdens while relocating complexity to policy configuration, verification, cross-component coordination, and workflow integration.
Our study provides useful lessons for the continuous design of identity-based software signing tools, particularly the need to treat verification semantics, policy validation, cross-component observability, and deployment integration as first-class usability targets rather than peripheral concerns.

\end{abstract}



\maketitle

\section{Introduction}
Modern software supply chains depend on organizations consuming code produced by others~~\cite{ossra-2020,okafor_sok}.
This reuse introduces attack vectors such as unauthorized or malicious code injection~~\cite{willett2023lessons, OSSRA_2024, benthall_assessing_2017, williams2025research, ladisa_sok_2023}.
To mitigate this risk, artifacts must carry verifiable \textit{provenance}, that is, evidence of integrity and origin, across organizational boundaries~~\cite{ladisa_sok_2023, kalu2025softwaresigningstillmatters, schorlemmer_establishing_2024}.
Software signing provides a cryptographically grounded mechanism for establishing provenance by binding artifacts to issuer identities and enabling downstream verification~~\cite{kalu2025softwaresigningstillmatters}.
Signing involves two coupled workflows: upstream signature creation by producers, and downstream signature verification by consumers.
If either workflow is hard to correctly complete, cryptographic guarantees degrade as producers mis-sign or consumers fail to verify~~\cite{schorlemmer_signing_2024}.

Usability challenges in software signing undermine its guarantees.
Early usable-security research showed that users struggled to correctly generate, manage, and verify cryptographic keys~~\cite{whitten1999johnny}.
Engineers adopt insecure workarounds when security workflows are complex~~\cite{green_developers_2016}.
Empirical studies of software signing ecosystems further document low adoption and persistent key-management failures in practice~~\cite{schorlemmer_signing_2024}.
In response to these burdens, modern ecosystems have increasingly adopted \textit{identity-based signing}, which replaces developer-managed long-lived keys with short-lived credentials issued by external identity and certificate services.
Empirical evidence on the usability of identity-based signing remains limited and is based primarily on interview studies~~\cite{kalu2025johnnysignsnextgenerationtools, usenix_2025_signing_interview_kalu}. 
These studies suggest that identity-based designs shift usability burdens by introducing new configuration and coordination challenges across tool components. 
However, we lack quantitative and longitudinal evidence about how reported usability concerns manifest across tools and change over time in modern signing ecosystems.
Longitudinal quantitative analysis can reveal whether reported usability concerns decline, remain relatively stable, or increase over time, and whether these trajectories differ across tools and problem types.
Such evidence can help identify usability concerns that persist or emerge as tools develop and are integrated into broader workflows, which is particularly important given that persistent usability burdens in key-managed signing coincided with low signing adoption outside mandatory settings~\cite{merrill2023speranza,schorlemmer_signing_2024,woodruff2023pgp_pypi_worse_than_useless}.

To provide cross-tool and longitudinal evidence on these usability challenges, we conduct the first mining-software-repositories investigation of usability issues in identity-based software signing tools.
We examine five prominent identity-based signing ecosystems and analyze user-reported usability concerns as reflected in GitHub issue discussions.
We code these discussions for the type of usability concern raised and the affected surface where the concern manifests, and we compare these patterns across tools.
For the longitudinal analysis, we draw on the Tool Usability Lifecycle, a conceptual framework that relates broader adoption and continued development to changing patterns of reported usability concerns.
We then use Poisson regression to characterize whether those reported concerns increase, remain relatively stable, or decline over time.
Together, this approach allows us to examine what usability problems are reported, where they manifest across the signing ecosystem, and how their reported frequency changes over the observation period.

Our results provide a cross-tool map of user-reported usability problems in identity-based software signing and an affected-surface view of where these problems arise.
Across ecosystems, CLI tooling, policy/configuration, and verification workflows are among the most frequently implicated affected surfaces.
Using longitudinal Poisson trend analysis, we find that reported issue counts decline significantly for several repositories, increase for OpenPubKey, and show no statistically detectable overall trend for the two Keyfactor repositories.
Individual themes and affected surfaces also follow different trajectories, indicating that reported usability concerns do not follow a uniform temporal pattern across identity-based signing ecosystems.
Together, these results show that identity-based designs shift usability challenges across components rather than eliminating them, and they identify concrete engineering surfaces where improvements are likely to strengthen real-world signing and verification guarantees.

To summarize our contributions:
\begin{itemize}[leftmargin=*, labelsep=0.5em, itemsep=1pt, topsep=1pt]
    \item We present the first repository-mining study of identity-based software-signing usability, providing cross-ecosystem evidence from five signing ecosystems.
    \item We provide an architecture-grounded characterization of user-reported usability concerns, mapping recurring themes to components and integration boundaries.
    \item We characterize how reported usability concerns change over time across tools, themes, and affected surfaces, revealing heterogeneous trends that inform targeted engineering improvements.
\end{itemize}

\paragraph{Study Significance}

Our work studies the usability of identity-based software signing tools, a class of signing systems that reduce the key-management burden of traditional signing by distributing signing responsibilities across interacting components.
These tools are increasingly integrated into software engineering workflows for build, release, and deployment, making their usability consequential to the practical adoption and operation of software supply-chain security mechanisms.
Our results show that usability challenges in these systems concentrate not only within individual interaction surfaces, but also at the boundaries between components and where signing infrastructure integrates with software delivery workflows.
This finding suggests that changes intended to simplify one part of a security workflow can redistribute usability friction across the surrounding software-engineering system.
Consequently, evaluating and designing identity-based signing requires attention to cross-component coordination and workflow integration, rather than considering individual interfaces in isolation.
Our findings specifically motivate semantically actionable status and error feedback, cross-component observability, and testable policy and configuration mechanisms as engineering priorities.
More broadly, these results provide evidence for designing and evaluating other security infrastructures composed of interacting services that must integrate with complex software engineering workflows.

\section{Background and Related Work}
\label{sec:Background}

  This section first introduces our usability framing, including how usability may change as tools evolve and see broader adoption (\cref{sec:Background:UsabilityTheory}), and then describes identity-based software signing and its architectural constituents (\cref{sec:Background_IdentitySigning}).
We finally define usability for this class of tools and position our work relative to prior studies of signing and adjacent security technologies (\cref{sec:Background:conceptual_usability_identitySigning}).

\subsection{Usability and Its Role in Tool Adoption and Maturity}
\label{sec:Background:UsabilityTheory}

\paragraph{Usability}  Usability concerns \textit{the extent to which a tool enables users to achieve its intended purpose}~~\cite{rusu2015user, lewis2014usability}.
In usability research and practice, usability assessments are of two kinds: summative and formative. These two kinds differ primarily in the actionability of their findings~~\cite{chapanis1981evaluating,lewis2014usability,theofanos_usability}.
Summative assessments measure the overall effectiveness of a design with respect to task completion~~\cite{lewis2014usability,iso9241-11,rusu2015user}, while formative assessments diagnose user challenges and identify opportunities for improvement~~\cite{theofanos_usability,lewis2014usability}.
Our study uses the summative lens to define usability-relevant barriers and follows the formative lens to diagnose where developers encounter those barriers in real signing, configuration, integration, and maintenance workflows (\cref{sec:method_human_annotation}).

\paragraph{Usability and Its Effect on Tool Adoption} The usability of a tool influences its adoption and acceptance~\cite{taherdoost2024navigating,lah2020perceived}.
Throughout this paper, we use \emph{usability friction} to refer to user-facing barriers that impede effective, efficient, or satisfactory completion of an intended task.
For software engineering tools, such friction can affect whether developers can integrate a tool into routine build, release, and deployment work.
This is especially important for security tools, where poor usability can lead to avoidance, misconfiguration, or incomplete use even when the underlying security mechanism is sound~\cite{nachtigall2022large,smith2020can}.
Because adoption exposes tools to broader users, new deployment environments, and workflow expectations, usability should be understood as a property that can change as tools evolve and are more broadly adopted.

\paragraph{Relationship Between Tool Adoption, Maturity, and Usability.}
Prior work suggests that usability needs and design priorities change as technologies evolve and reach broader user populations.
Technology adoption models describe diffusion from early adopters toward broader use~\cite{rogers2014diffusion, moore1991crossing}, while technology-maturity models characterize how technologies develop and stabilize over time~\cite{nolte_2008}.
Complementing these adoption and maturity perspectives, usability engineering emphasizes iterative evaluation and refinement as systems encounter user needs and usability problems~\cite{Nielsen_1993}.
Subramanian brings these perspectives together in a Design Strategy Framework that relates technology evolution and market adoption to changing design priorities, including usability~\cite{subramanian2009dynamically}.
However, this framework was developed for consumer technologies, such as smartphones, and does not specifically characterize how user-reported usability concerns evolve in software tools.

We therefore extend this lifecycle perspective to software tools as the \emph{Tool Usability Lifecycle}, a conceptual framework that relates broader adoption and continued development to changes in reported usability concerns.
~\cref{tab:UsabilityLifecycle} summarizes the lifecycle stages and the usability signals associated with them. ~\cref{fig:tool-usability-lifecycle} illustrates the mechanisms underlying these signals: broader adoption may expand exposure to users and deployment contexts, while continued development may enable iterative remediation of recurring concerns.
The balance between these processes may produce increasing, stable, or declining reported concerns, while the prominence of concerns may also shift across affected surfaces.
We use the Tool Usability Lifecycle as a conceptual lens rather than as a framework that prescribes a single temporal trajectory.
Specifically, it identifies expanding exposure and iterative remediation as competing mechanisms that may produce increasing, relatively stable, or declining patterns of reported usability concerns, without assuming that all tools follow the same trajectory.
This lifecycle framing motivates RQ3's exploratory longitudinal analysis (\cref{sec:rqs}), which characterizes how user-reported usability concerns change across tools and affected surfaces.

\begin{table}[!t]
\centering
\caption{
  The \emph{Tool Usability Lifecycle}, a conceptual framework synthesizing prior work on usability, technology adoption, and technology maturity~\cite{Nielsen_1993,moore1991crossing,nolte_2008,rogers2014diffusion,subramanian2009dynamically}.
  Because these prior perspectives address related but distinct aspects of technology evolution, the framework brings them together to characterize how user-reported usability concerns may rise, shift, stabilize, or decline as software tools experience broader adoption, integration, and continued development.
}

\small
\footnotesize
\label{tab:UsabilityLifecycle}
\begin{tabular}{p{0.20\linewidth} p{0.30\linewidth} p{0.38\linewidth}}
\toprule
\textbf{Lifecycle Stage} & \textbf{Signals} & \textbf{Mechanisms} \\
\midrule
Early Adoption & Rising reports of friction and configuration issues & Increased exposure; incomplete documentation; unstable workflows \\
Growth / Integration & Reported concerns remain frequent or shift across problem types and affected surfaces & Integration into CI/CD; policy refinement; identity management complexity \\
Continued Development / Stabilization & Stabilization or decline in issue frequency & Improved defaults; community knowledge; workflow automation \\
\bottomrule
\end{tabular}
\end{table}

\FloatBarrier

\begin{figure*}[!t]
    \centering
    \includegraphics[width=0.87\textwidth,trim=21bp 2512bp 5080bp 50bp,clip]{fig/new/tool-usability-lifecycle_e.pdf}
\caption{The \emph{Tool Usability Lifecycle} illustrates how expanding exposure and iterative remediation may shape user-reported usability concerns over time.
The framework is a conceptual lens rather than a prescribed temporal curve; a fixed observation window may capture only part of a tool's broader lifecycle.}
    \label{fig:tool-usability-lifecycle}
\end{figure*}

\subsection{Identity-Based Software Signing}
\label{sec:Background_IdentitySigning}

Software signing uses public-key cryptography to establish artifact integrity and signer authenticity or provenance~~\cite{schorlemmer2025establishing}.
Traditional key-managed approaches (\eg PGP/GPG) proved difficult to use in practice due to key lifecycle management~~\cite{whitten1999johnny,Garfinkel2003-pa}.
Recently, identity-based signing systems were developed --- these reduce key-management friction via short-lived certificates bound to federated identities~~\cite{newman_sigstore_2022,merrill2023speranza}.
We review the concepts of identity-based signing (\cref{sec:Background:IdentitySigning:Concepts}) and then situate our study by summarizing prior empirical findings on signing usability (\cref{sec:Background:IdentitySigning_PriorUsability}).

\begin{table*}[t]
\centering
\caption{Core components of identity-based signing ecosystems and their responsibilities.}
\label{tab:id_components}
\small
\scriptsize
\footnotesize
\begin{tabular}{p{0.28\textwidth} p{0.68\textwidth}}
\toprule
\textbf{Component} & \textbf{Primary Responsibilities and Typical Implementations} \\
\midrule
\textbf{Signing workflow coordinator} &
Primary developer interface for signing and verification.
Creates or selects artifacts, obtains identity tokens, generates ephemeral key pairs, creates signatures, and associates signatures with credential material.
Examples include Sigstore \texttt{cosign}~~\cite{sigstore_website, sigstore_cosign_github}, Notary~v2 \texttt{notation}~~\cite{notaryproject_website, notaryproject_notation_github}, and OpenPubKey tooling~~\cite{openpubkey_github, heilman_openpubkey_2023}. \\
\addlinespace
\textbf{Identity Provider (IdP)} &
Authenticates a human or workload and issues an identity token presented to other services.
Often implemented using standards-based systems such as OpenID Connect providers.
Examples include enterprise IdPs and developer-ecosystem identities (\eg Google or GitHub-based authentication), depending on configuration and deployment model. \\
\addlinespace
\textbf{Credential Issuer / CA} &
Binds authenticated identity to signing material, commonly by issuing a short-lived certificate for an ephemeral public key.
Enables time-scoped and auditable identity binding.
Examples include Sigstore deployments using Fulcio~~\cite{sigstore_fulcio_github} and enterprise PKI systems such as EJBCA~~\cite{ejbca_ce_github, ejbca_website}. \\
\addlinespace
\textbf{Certificate and Signature Logs} &
Append-only logs that record certificate issuance and signing events to support transparency, monitoring, and audit.
Shape verification semantics and operational practices.
Example: Sigstore’s Rekor~~\cite{sigstore_rekor_github}. \\
\bottomrule
\end{tabular}
\end{table*}

\subsubsection{Concepts and Architecture}
\label{sec:Background:IdentitySigning:Concepts}
Software signing is the primary cryptographic mechanism used to establish artifact integrity and provenance in modern software supply chains~~\cite{kalu2025johnnysignsnextgenerationtools, kalu2025softwaresigningstillmatters, Garfinkel2003-pa, rfc4880, Internet-Security-Glossary, cooper_security_2018}.
It builds on public key cryptography, where a signer holds a private key and distributes a corresponding public key, and a verifier verifies the identity of the signer using the distributed public key~~\cite{diffie1976new, katzlindell2014crypto, schorlemmer_establishing_2024}.
When verification succeeds, the verifier gains evidence that the artifact has not been modified since signing and that the signer controlled the corresponding private key at signing time~~\cite{katzlindell2014crypto}.
In practice, these guarantees depend not only on cryptographic primitives but also on how signing keys or credentials are created, protected, rotated, revoked, discovered, and bound to identities~~\cite{whitten1999johnny, kalu2025johnnysignsnextgenerationtools, usenix_2025_signing_interview_kalu, schorlemmer_signing_2024}.
Conventional key-managed signing workflows, described in~\cref{sec:bg_trad_signing}, require developers to create and manage long-lived keys directly. This responsibility has historically posed persistent usability and operational challenges~~\cite{whitten1999johnny}.

Identity-based signing is a recent response to this challenge.
This approach replaces developer-managed long-lived keys with short-lived credentials issued by external identity and certificate services~~\cite{okafor2024diverify, schorlemmer_establishing_2024}.
Instead of requiring developers to create and distribute long-lived keys, identity-based ecosystems authenticate an actor through an identity provider and bind that identity to signing material through a credential-issuing service.
This approach commonly issues short-lived certificates for ephemeral public keys, enabling verification to rely on time-scoped and auditable identity bindings rather than indefinitely trusted keys.
As a result, identity binding and credential lifecycle management are shifted from individual actors to shared ecosystem infrastructure.

Identity-based signing ecosystems are typically architected using four primary components. ~\cref{tab:id_components} elaborates on each component. Succinctly:


\begin{figure*}
    \centering
    \includegraphics[width=0.850\linewidth,trim=35bp 467bp 30bp 85bp,clip]{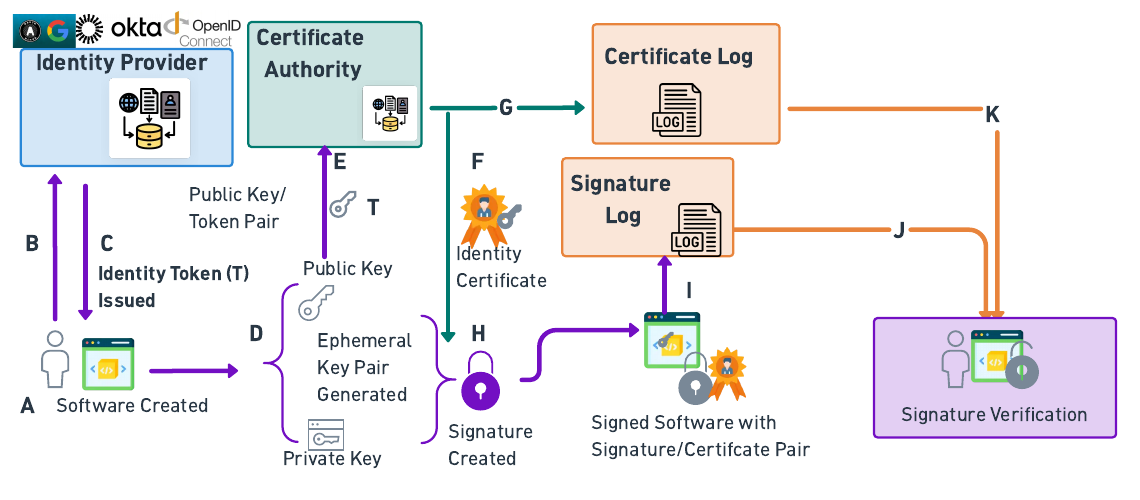}
    \caption{
    Identity-based signing uses short-lived certificates instead of long-lived keys.
    Typical architectures decompose signing and verification across four components (\cref{tab:id_components}). 
Purple steps (ABCDEHI) show actions coordinated by the \emph{Signing Workflow Coordinator}, including identity-token retrieval, ephemeral key generation, signature creation, and submission of signing evidence to logs.
Green steps (F--G) show \emph{credential-issuer/Certificate Authority} actions that bind the authenticated identity to signing material by issuing a short-lived certificate.
Orange steps (J--K) show \emph{certificate and signature log} actions that record evidence used during verification.
Blue indicates the identity provider, which authenticates the signer and issues the identity token used in the signing workflow.
Here, $T$ denotes the identity token issued by the identity provider; it is paired with the ephemeral public key to obtain the short-lived identity certificate.
    }
    \label{fig:bg_id_workflow}
\end{figure*}

\begin{itemize}[leftmargin=*, labelsep=0.5em, itemsep=1pt, topsep=1.5pt]
\item \textbf{Signing workflow coordinator:}
Developer-facing component that creates or selects artifacts, obtains identity tokens, generates ephemeral key pairs, produces signatures over artifacts, and coordinates with other signing-system components during signing or verification.

\item \textbf{Identity Provider (IdP):}
Authenticates a human or workload and issues an identity token for other services.

\item \textbf{Credential Issuer / Certificate Authority:}
Binds authenticated identity to signing material, commonly by issuing a short-lived certificate for an ephemeral public key.

\item \textbf{Certificate and Signature Logs:}
Append-only logs that record certificate issuance and signing events to support transparency, monitoring, and audit.
\end{itemize}

\noindent
These components collectively implement the workflow illustrated in~\cref{fig:bg_id_workflow}.
Compared with the key-managed signing workflow shown in~\cref{fig:bg_trad_workflow}, this architectural decomposition allows components and services to be reused and customized across ecosystems.
At the same time, distributing signing across additional services and interfaces creates more configuration points and integration boundaries.
These surfaces and boundaries shape how signing and verification are configured and performed, and therefore where users may encounter usability issues.

\subsection{Usability in Identity-Based Signing}
\label{sec:Background:conceptual_usability_identitySigning}
Identity-based signing tools are architecturally complex sociotechnical systems rather than discrete user-interface applications~\cite{usenix_2025_signing_interview_kalu}.
They are operated through interacting infrastructure components described in \cref{sec:Background:IdentitySigning:Concepts} and are exposed to users primarily through CLI commands, API calls, configuration files, and verification output rather than a single graphical interface~\cite{kalu2025johnnysignsnextgenerationtools,usenix_2025_signing_interview_kalu}.
Empirical studies characterize software signing as involving interactions among users, organizational practices, and the technical infrastructure through which signing is integrated into development and release workflows~\cite{kalu2025johnnysignsnextgenerationtools,usenix_2025_signing_interview_kalu}.
These characteristics are typical of complex systems, in which usability emerges not only from individual interfaces but also from interactions among components, environmental conditions, and the situated workflows in which the system is used~\cite{Redish_2007,Harrison,Mirel,Pan_Komandur_Finken_2015,Norros_Savioja}.
Usability concerns may therefore arise both within individual components and at the boundaries between them, where configuration, feedback, documentation, or workflow-integration problems can impede successful signing and verification.


Prior software-failure studies have emphasized characterizing not only reported problems but also their manifestation and the components they affect~\cite{amusuo2022reflections}.
Because usability in these complex systems extends beyond individual interfaces, our analysis must account for where within the broader signing ecosystem a reported concern manifests.
Throughout this paper, we use \emph{affected surface} to denote the component, interface, workflow, or integration boundary at which a user-reported usability concern manifests.
This construct connects the architectural decomposition summarized in~\cref{tab:id_components} to our empirical analysis and is operationalized as part of the issue-coding scheme in~\cref{sec:method_processing}.

\subsubsection{Empirical Evidence on Signing Usability}
\label{sec:Background:IdentitySigning_PriorUsability}

Usability challenges in software signing were first documented in studies of key-managed cryptographic tools like communication encryption, which showed that public-key workflows impose substantial cognitive and operational burden on users.
Early usable-security research demonstrated that ordinary computer users like ``Johnny,'' who were novices in cryptography, struggled to correctly generate, manage, and verify cryptographic keys, often misunderstanding trust models, misconfiguring tools, and failing to execute key lifecycle tasks correctly~~\cite{whitten1999johnny, sheng2006johnny, ruoti_confused_2013, reuter_secure_2020}. 
Even twenty years later,  empirical measurements of software package registries showed low signing rates outside mandatory settings, and that public-key management failures are common in maintainer-managed workflows~~\cite{merrill2023speranza,schorlemmer_signing_2024}.
In key-based signing, signing failures are due to usability issues more than faulty cryptography~\cite{schorlemmer2025establishing}.

Recent work has begun to examine the usability of identity-based signing.
To date, only interview-based data is available about the usability of identity-based signing~~\cite{kalu2025johnnysignsnextgenerationtools, usenix_2025_signing_interview_kalu}.
Interviews indicate that in practice, identity-based approaches alter but do not eliminate the usability situation. 
Identity-based signing introduces new configuration complexity and coordination challenges, and redistributes workflow roles across signing workflow coordinators, identity providers, and credential issuers.
These studies suggest that identity-based designs change where usability problems emerge rather than eliminating them.

\subsection{Summary of Knowledge Gaps.}

Prior work establishes that software-signing usability remains challenging and that identity-based approaches can introduce configuration, integration, and coordination difficulties.
However, three important knowledge gaps remain.
These gaps matter because the practical value of signing tools depends partly on whether users can complete security-sensitive tasks without excessive usability friction.
Addressing them can clarify which usability problems recur across ecosystems, where they manifest within signing workflows, and how they change over time, providing evidence that can help maintainers prioritize usability improvements.
\begin{itemize}[leftmargin=*, labelsep=0.5em, itemsep=1pt, topsep=1pt]

    \item \textit{Limited comparative evidence across ecosystems.}
    Existing evidence on identity-based signing usability is based primarily on interview studies, leaving limited cross-ecosystem quantitative evidence about the usability concerns reported in practice.
    This limits our ability to distinguish broadly recurring concerns from those that are specific to particular tools or ecosystems.

    \item \textit{Limited evidence about where usability concerns manifest.}
    Prior work does not systematically characterize where these concerns arise across the components, interfaces, workflows, and integration boundaries of identity-based signing ecosystems.
    Identifying these affected surfaces can help reveal where usability barriers interfere with signing, verification, integration, and maintenance tasks, and where engineering interventions may be most useful.

    \item \textit{Limited longitudinal evidence.}
    We lack evidence about how reported usability concerns change over time, including whether different tools, concern types, and affected surfaces follow similar or divergent trajectories.
    Without such evidence, it is difficult to distinguish concerns that remain prominent from those that decline or emerge as tools continue to develop and are integrated into broader workflows.

    These gaps motivate our research questions in the following section.

\end{itemize}

\section{Research Questions}
\label{sec:rqs}

We address the three knowledge gaps identified above through the following research questions:
\begin{itemize}[leftmargin=*, labelsep=0.5em, itemsep=1pt, topsep=1pt]
\item \textit{\textbf{RQ1}: What usability problems are reported for identity-based software signing tools?}
\item \textit{\textbf{RQ2}: Which affected surfaces of identity-based software signing tools are most frequently involved in usability problems?}
\item \textit{\textbf{RQ3}: How do reported usability problems change over time across identity-based software signing tools?}

RQ3 is exploratory and is guided by the \emph{Tool Usability Lifecycle} introduced in \cref{sec:Background:UsabilityTheory}.
Rather than assuming a uniform temporal trajectory, we characterize whether user-reported usability concerns increase, remain stable, or decline over the observation period and compare these patterns across tools and affected surfaces.
\end{itemize}


\section{Methodology}
\label{sec:method}

\begin{figure*}
    \centering
    \begin{minipage}{0.9\linewidth}
        \noindent
        \includegraphics[page=1,width=0.64010\linewidth,trim=244bp 27bp 18bp 1095bp,clip]{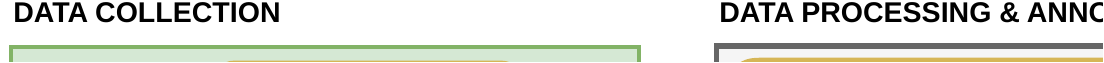}%
        \hspace{-0.0035\linewidth}%
        \includegraphics[page=2,width=0.35990\linewidth,trim=18bp 27bp 476bp 1095bp,clip]{fig/methodology_revised_a.pdf}%
        \par\nointerlineskip
        \noindent
        \includegraphics[page=3,width=0.64010\linewidth,trim=244bp 574bp 18bp 27bp,clip]{fig/methodology_revised_a.pdf}%
        \hspace{-0.0035\linewidth}%
        \includegraphics[page=4,width=0.35990\linewidth,trim=18bp 574bp 476bp 27bp,clip]{fig/methodology_revised_a.pdf}%
    \end{minipage}
    \caption{Overview of our methodology for analyzing user-reported usability concerns in identity-based software signing tools using GitHub issues.
The pipeline shows data collection, stratified human-sample design, human coding-scheme development using inductive codebook-based thematic analysis and deductive Nielsen heuristic mapping, held-out validation of LLM-assisted coding, and full-corpus annotation.
The resulting annotations support RQ1 (usability theme characterization), RQ2 (affected-surface distribution), and RQ3 (exploratory longitudinal trend analysis).}
    \label{fig:method_overview}
\end{figure*}

Our methodology comprises three stages, summarized in Figure~\ref{fig:method_overview}: (1) data collection, (2) data processing and annotation, and (3) data analysis.
First, we select identity-based signing ecosystems and collect and filter their GitHub issue threads (~\cref{sec:method_tool_selection_data_collection}).
Second, we construct the human sample, develop and validate the coding scheme, and scale annotation to the full corpus (~\cref{sec:method_processing}).
Third, we analyze the resulting corpus to characterize usability themes (RQ1), affected-surface distributions (RQ2), and exploratory longitudinal trends in user-reported usability concerns (RQ3) (~\cref{sec:analysis}).

\subsection{Data Collection}
\label{sec:method_tool_selection_data_collection}

Here, we justify our choice of identity-based signing tools (\cref{sec:method_tools}), the suitability of our data source (\cref{sec:method_issues_justification}), and our data collection process (\cref{sec:method_data_collection}).

\subsubsection{Identity-Based Signing Tool Selection}
\label{sec:method_tools}

To be included in our study, a tool must fully or partially follow the identity-based workflow described in \cref{sec:Background:IdentitySigning:Concepts}.

We identify prominent identity-based signing ecosystems based on their presence in the literature and documented use in practice~\cite{okafor2024diverify, usenix_2025_signing_interview_kalu, kalu2025johnnysignsnextgenerationtools, schorlemmer2025establishing}.
We also consider the availability of publicly documented evidence of practical adoption, including academic studies, practitioner reports, engineering blogs, and project or vendor case studies.\footnote{We also considered managed signing services such as AWS Signer~\cite{aws2024signer} and Microsoft Artifact Signing~\cite{microsoft_artifact_signing_overview}.
While these services are relevant to the broader software-signing landscape, we found comparatively less publicly documented evidence of adoption in the academic literature, practitioner reports, and public case studies we reviewed than for the ecosystems selected for this study.}

This process yields five identity-based signing ecosystems:
\textit{Sigstore}~\cite{sigstore_dev} (used by PyPI~\cite{Ingram2024PyPI}, Yahoo~\cite{yahoo_scaling_supply_chain}, Kubernetes~\cite{Vaughan-Nichols_2022_Kubernetes}, etc.),
\textit{Notary~v2}~\cite{notaryproject_notation_github} (used by AWS and Microsoft~\cite{notaryproject_website}),
\textit{OpenPubKey}~\cite{openpubkey_github} (used by Docker~\cite{stoten2023_signing_docker_official_images_openpubkey}, BastionZero~\cite{bastionzero_website}),
\textit{Keyfactor}~\cite{keyfactor_website} (users include Siemens and Schneider Electric~\cite{keyfactor_customers}),
and IBM's \textit{HashiCorp Vault}, used by Mantech and others~\cite{hashicorp_vault_web,ibm2025_hashicorp_acquisition_complete,hashicorp_mantech_case_study,hashicorp_vault_product}.

Notably, all five selected ecosystems expose public development repositories and issue discussions.
This public development record makes it possible to study reported usability concerns longitudinally using repository-mining methods.

We only select components of these tools corresponding to the architecture described in \cref{fig:bg_id_workflow}.
For example, for Sigstore, we include the signing workflow coordinator (Cosign), the certificate authority (Fulcio), and the log (Rekor).
Our component selection was based on analysis of official documentation, repository structure, and workflow descriptions for each ecosystem, and was validated through author consensus.
\cref{tab:tools_components} summarizes the repositories, their component roles, and the number of issues analyzed per component.


\begin{table*}[t]
\centering
\caption{
The analyzed projects and their organizational and temporal context.
\emph{Documented Start} denotes the earliest documented public or operational start of the analyzed ecosystem or component.
These dates provide descriptive context for the histories of the analyzed tools and are not treated as measures of maturity.
}
\scriptsize
\footnotesize
\setlength{\tabcolsep}{2.8pt}
\renewcommand{\arraystretch}{1.15}
\begin{tabularx}{\textwidth}{@{}l l >{\hsize=1.15\hsize\linewidth=\hsize\raggedright\arraybackslash}X >{\hsize=0.85\hsize\linewidth=\hsize\raggedright\arraybackslash}X c >{\centering\arraybackslash}p{0.085\textwidth} r@{}}
\toprule
\textbf{Tool} & \textbf{Component} & \textbf{Role} & \makecell[l]{\textbf{Sponsor/}\\\textbf{Host}} & \makecell{\textbf{Documented}\\\textbf{Start}} & \makecell{\textbf{Stars}\\\textbf{(Forks)}} & \makecell{\textbf{Issues}\\\textbf{($n$)}} \\
\midrule
Sigstore   & Cosign      & Signing workflow coordinator          & OpenSSF / Linux Fdn. & Mar.~2021 & 5.6K (687)  & 871 \\
 & Fulcio      & Certificate authority                & OpenSSF / Linux Fdn. & Mar.~2021 & 796 (168)   & 196 \\
 & Rekor       & Certificate/signature log             & OpenSSF / Linux Fdn. & Mar.~2021 & 1.1K (197)  & 251 \\
\midrule
Keyfactor  & SignServer  & Signing workflow coordinator          & Keyfactor & 2005 & 407 (51)    & 23 \\
 & EJBCA-CE    & Certificate authority               & Keyfactor & 2001 & 869 (155)   & 189 \\
\midrule
Notary~v2  & Notation    & Signing workflow coordinator          & CNCF / Linux Fdn. & Dec.~2019 & 460 (91)    & 333 \\
OpenPubKey & OpenPubKey  & Signing workflow coordinator          & Linux Fdn. & Early 2021 & 881 (68)    & 147 \\
HashiCorp  & Vault       & Main interface / credential issuer    & HashiCorp / IBM & Apr.~2015 & 35K (4.6K)  & 1890 \\
\bottomrule
\end{tabularx}
\vspace{2pt}
\parbox{\textwidth}{\scriptsize\textit{Notes.}$n$ = number of issues analyzed; Fdn. = Foundation. For Notary~v2, the broader Notary project originated earlier; December~2019 denotes the start of the Notation implementation we analyzed. For OpenPubKey, early~2021 reflects the underlying BastionZero implementation already being in operational use, rather than the October~2023 launch of OpenPubKey as a standalone Linux Foundation project.}
\label{tab:tools_components}
\end{table*}

\subsubsection{Data Source: GitHub Issues as Usability Evidence}
\label{sec:method_issues_justification}

We use GitHub issues and their associated comment threads to study user-reported usability problems in identity-based software signing tools.
Issue reports are a natural data source for this study because they capture problems encountered during real adoption, integration, and maintenance work, and they preserve the context of troubleshooting and resolution through discussion~~\cite{cheng_how_CHI_2018}.
This source is also aligned with our goal of characterizing usability problems that arise in practice, rather than usability judgments elicited through controlled tasks.
As is standard in mining software repositories research, we treat issue discussions as situated evidence of developer experience rather than as a complete record of all problems encountered~~\cite{kalliamvakou2014promises}.
Prior work has likewise used GitHub issues and other developer discussion platforms, such as Stack Overflow, as sources of usability evidence~~\cite{Sanei_Cheng_2024, cheng_how_CHI_2018, Patnaik_Hallett_Rashid, Yang_Wang_Shi_Hoang_Kochhar_Lu_Xing_Lo_2023}.


\subsubsection{Issue Collection}
\label{sec:method_data_collection}
We collect GitHub issues from the selected repositories from November~2021 through November~2025, yielding 49 inclusive calendar-month bins over 48 elapsed months.
We choose this window to enable longitudinal analysis while keeping the dataset within a fixed time horizon in which tool workflows and developer platforms are broadly comparable.
To keep the comparison contemporaneous across tools, we use this fixed recent window rather than sampling across each repository's full history.
For each issue, we retain issue metadata and link the issue to its comment thread to preserve troubleshooting and resolution context.
We exclude bot-generated issues to focus our analysis on user-reported usability problems.
In total, our corpus contains approximately 3,900 issues across the eight selected component repositories in the five ecosystems, with per-repository counts summarized in \cref{tab:tools_components}.
Appendix~\cref{tab:issue-context-distributions-by-repo} reports per-repository issue-thread context metadata, including comments, participants, and approximate text length.

\subsection{Data Processing and Annotation}
\label{sec:method_processing}


This section describes how we develop the human coding scheme, validate its LLM-assisted application, and scale annotation to the full corpus.
Our data-processing and annotation procedure is structured by complementary qualitative and usability-analysis approaches that guide the sequence of coding and scaling steps summarized in the \emph{Data Processing and Annotation} stage of~\cref{fig:method_overview}.

For the inductive analysis, we draw on Guest \etal's codebook-based thematic analysis~\cite{guest2011applied}, which emphasizes familiarization with the data, systematic development and refinement of codes and themes, and consistent application of an explicit coding scheme.
This familiarization was particularly important because usability evaluation in complex domain-specific systems can depend on sufficient understanding of the underlying domain and workflows~\cite{chilana_understanding_2010}.
Accordingly, our human analysis proceeds from familiarization and shared understanding to usability-relevance coding and inductive theme development, culminating in a finalized coding scheme and annotation rules.
We complement this inductive analysis with a deductive framework analysis approach~\cite{srivastava2009framework}, using Nielsen's usability heuristics as the predefined analytical framework~\cite{nielsen_10_usability_heuristics}.
Together, these complementary approaches allow concerns specific to identity-based signing to emerge from the issue discussions while also supporting systematic comparison across tools using an established usability vocabulary.

To extend the human-developed scheme beyond the development sample, we draw on prior guidance for LLM-assisted annotation and qualitative coding~\cite{Alizadeh_Kubli_Samei_Dehghani_Zahedivafa_Bermeo_Korobeynikova_Gilardi_2024,wang2024human,peng2025agentsperformcodeoptimization}.
Our procedure uses a human-coded development set to establish and refine the annotation instructions, applies the resulting scheme with the LLM, and evaluates its annotations against a separately held-out human-coded set before full-corpus annotation.
The following subsections describe the human coding-scheme development and LLM-assisted scaling and validation stages in detail.

\subsubsection{Establishing an Issue Annotation Scheme: Human Pair Annotation}
\label{sec:method_human_annotation}

We began issue processing by developing a coding scheme for issue content, usability relevance, affected surfaces, and usability themes.
Two analysts (referred to as A and B) conducted this phase.\footnote{Both analysts had experience in qualitative coding and HCI research, while Analyst~A additionally had prior domain knowledge of identity-based signing.
The calibration step described in ~\cref{para:method_calibrating_issue_understanding} was designed to establish a shared understanding of identity-based signing concepts and issue context before usability classification.}

We selected a stratified human-annotation sample of 180 issues to support two distinct methodological purposes: development of the coding scheme and independent validation of its subsequent LLM-assisted application.
We allocated 100 issues to the development set to provide a sufficiently broad human-coded basis for familiarization, usability-relevance classification, theme development, deductive heuristic mapping, and refinement of the annotation instructions.
This development-set size is consistent with prior LLM-assisted text-annotation work that evaluates the use of human-labeled development sets on the order of tens to hundreds of examples, including 50 and 100 example conditions~\cite{Alizadeh_Kubli_Samei_Dehghani_Zahedivafa_Bermeo_Korobeynikova_Gilardi_2024}.
We reserved the remaining 80 issues as a held-out human-coded validation set that was not used during coding-scheme or annotation-instruction development.
The held-out set provides an independent basis for evaluating the LLM's application of the finalized scheme before full-corpus annotation.
For proportion-based validation measures, a sample of 80 yields a worst-case 95\% margin of error of approximately $\pm 11$ percentage points~\cite{cochran1977sampling}\footnote{Using the normal approximation for a proportion with finite-population correction, the worst case occurs at $p=0.5$: $1.96\sqrt{0.5(1-0.5)/80}\sqrt{(3900-80)/(3900-1)}\approx0.1085$, corresponding to a margin of error $\approx$ $\pm 10.9$ percentage points.}, providing a meaningful estimate of agreement while retaining sufficient human-coded data for scheme development.
Thus, the 180-issue sample reflects the combined requirements of coding-scheme development and independent validation rather than a target percentage of the full corpus.

We sampled the 180 issues proportionally across repositories and further stratified by time (early, middle, and late thirds of the observation window), discussion intensity (short versus long comment threads), and issue labels (\eg bug, enhancement, question, documentation) to expose the coding process to variation in issue context and reporting style.
We used repository labels to guide stratification rather than as ground-truth classifications.

\paragraph{Calibrating Issue Understanding Between Analysts.}
\label{para:method_calibrating_issue_understanding}

Before usability classification, the analysts established a shared understanding of identity-based signing concepts and the context represented in the issue threads.
Following qualitative-coding guidance recommending that analysts independently examine a subset of the data to calibrate interpretation before broader coding~\cite{oconnor_intercoder_2020,campbell2013coding}, both analysts independently reviewed and memoed 50 issues from the 100-issue development set.
We selected 50 issues to provide substantial exposure to the variation across repositories, time periods, discussion intensities, and issue types, particularly because both analysts did not share prior domain knowledge of identity-based signing.
For each issue, they reviewed the title, repository labels, body, and associated comments and recorded a plain-language description of the reported concern and the affected surface implicated by the discussion.
This was an open familiarization and calibration exercise, not a closed-label coding task.
The analysts compared their memos, discussed differences in interpretation, and refined their shared understanding of the signing workflows, components, and issue context before proceeding to usability classification.

\paragraph{Classifying Usability Versus Non-Usability Issues.}

After establishing a shared understanding of issue context, both analysts independently coded the 100-issue development set for usability relevance, classifying each issue as usability-relevant or non-usability.
We operationalized usability using the summative and formative perspectives introduced in \cref{sec:Background:UsabilityTheory}: summative usability assesses whether users can achieve specified goals effectively, efficiently, and satisfactorily, while formative usability diagnoses difficulties users encounter and identifies opportunities for improvement~\cite{lewis2014usability,theofanos_usability,iso9241-11,rusu2015user}.
For identity-based signing, these goals include signing and verifying artifacts, configuring identity and trust policies, integrating signing into release workflows, interpreting verification evidence, and maintaining the components that support these tasks.

Consistent with prior mining-software-repositories studies that characterize usability through issue discussions~\cite{Sanei_Cheng_2024,cheng_how_CHI_2018}, we treated an issue as usability-relevant when the discussion exposed a user-facing barrier to achieving an intended signing, verification, integration, or maintenance goal effectively, efficiently, or satisfactorily.
Such barriers could arise from workflow design, interface behavior, feedback, documentation, configuration, or missing workflow affordances.
When the initial report was ambiguous, the analysts considered the full issue thread, including follow-on discussion and the apparent resolution, rather than relying solely on the reporter's initial description or repository labels.
For example, a user may initially report a functional defect that subsequent discussion reveals to involve confusion about configuration or tool behavior.
When an issue contained both functional and usability aspects, we coded it as usability-relevant when the discussion exposed a user-facing workflow, feedback, documentation, configuration, or missing-affordance barrier beyond the underlying functional defect.
~\Cref{tab:method_usability_operationalization} summarizes this decision boundary with representative included and excluded issue patterns.

Initial percent agreement between the analysts was 70\%.
Following established guidance on iterative intercoder comparison and codebook refinement~\cite{oconnor_intercoder_2020}, the analysts discussed disagreements and refined the usability definitions and identification rules in the codebook.
All remaining disagreements were then resolved through discussion.

\begin{table*}[t]
\centering
\caption{Operational criteria and paired issue examples for distinguishing usability-relevant from non-usability issue discussions in identity-based signing. The criteria make explicit whether an issue exposes a user-facing barrier to task completion or understanding, while the paired examples illustrate how that boundary was applied during coding.}
\label{tab:method_usability_operationalization}
\scriptsize
\footnotesize
\setlength{\tabcolsep}{4.5pt}
\renewcommand{\arraystretch}{1.15}
\begin{tabularx}{\textwidth}{@{}p{0.17\textwidth} X X@{}}
\toprule
\textbf{Evidence category} & \textbf{Coded as usability-relevant when...} & \textbf{Coded as non-usability when...} \\
\midrule
Configuration and policy & \textit{Criterion:} The discussion shows difficulty specifying, understanding, or maintaining signing or verification configuration needed for a legitimate workflow.\newline\textit{Example:} \href{https://github.com/sigstore/fulcio/issues/1178}{Fulcio~\#1178} describes unclear AWS KMS URI syntax, IAM permissions, and certificate-chain configuration that force the reporter to inspect source code and experiment. & \textit{Criterion:} The issue concerns an internal policy-file refactor or default-setting cleanup with no observable effect on a user's signing, verification, integration, or maintenance task.\newline\textit{Example:} \href{https://github.com/notaryproject/notation/issues/1151}{Notation~\#1151} refactors internal command-output and input-flag handling without describing a change to such a task. \\
\midrule
Integration and workflow & \textit{Criterion:} A user cannot complete or sustain signing, verification, release, CI/CD, or component-coordination work because the tool workflow is unclear, incomplete, or difficult to operate.\newline\textit{Example:} \href{https://github.com/notaryproject/notation/issues/696}{Notation~\#696} reports that \texttt{notation login} fails under WSL when Docker uses the \texttt{desktop.exe} credential store, requiring manual Notation configuration. & \textit{Criterion:} The issue is limited to internal build infrastructure, repository maintenance, or implementation restructuring without a user-facing workflow consequence.\newline\textit{Example:} \href{https://github.com/sigstore/rekor/issues/2395}{Rekor~\#2395} moves an internal Terraform reference used by performance-testing infrastructure. \\
\midrule
Feedback and documentation & \textit{Criterion:} Error messages, verification output, status information, or documentation make it difficult to understand trust semantics, diagnose failures, or recover.\newline\textit{Examples:} \href{https://github.com/notaryproject/notation/issues/625}{Notation~\#625} requires debug output to reveal a certificate-identity mismatch hidden by a general error; \href{https://github.com/sigstore/cosign/issues/3671}{Cosign~\#3671} describes verification documentation that omits required identity and OIDC-issuer flags. & \textit{Criterion:} The issue is a typo, style edit, or documentation housekeeping item with no connection to task completion or user understanding.\newline\textit{Example:} \href{https://github.com/sigstore/fulcio/issues/1301}{Fulcio~\#1301} corrects a typo without a described task-level consequence. \\
\midrule
Workflow-level feature request & \textit{Criterion:} The requested feature or affordance is needed to complete, simplify, or make reliable a legitimate signing, verification, integration, or maintenance task.\newline\textit{Example:} \href{https://github.com/openpubkey/openpubkey/issues/301}{OpenPubKey~\#301} reports that users cannot use Okta or self-hosted OIDC providers because provider endpoints are not dynamically configurable. & \textit{Criterion:} The request is speculative, cosmetic, or unrelated to a concrete workflow barrier observed in the report or discussion.\newline\textit{Example:} \href{https://github.com/openpubkey/openpubkey/issues/321}{OpenPubKey~\#321} explores support for the OIDC \texttt{claims} parameter without describing a concrete current workflow that is blocked or difficult to complete. \\
\midrule
Functional defects and performance & \textit{Criterion:} A defect or slowdown creates a user-facing workflow barrier, misleading interface contract, unclear recovery path, or other obstacle beyond the internal fault itself.\newline\textit{Example:} \href{https://github.com/sigstore/fulcio/issues/1858}{Fulcio~\#1858} reports a mismatch between advertised RSA-PSS support and actual API behavior that causes signing or verification failures and misrepresents the supported workflow. & \textit{Criterion:} The issue is primarily a localized implementation defect, conformance failure, or internal performance or code-quality change with no user-facing workflow impact.\newline\textit{Example:} \href{https://github.com/openpubkey/openpubkey/issues/290}{OpenPubKey~\#290} re-enables linting and resolves internal findings after a Go upgrade. \\
\bottomrule
\end{tabularx}
\end{table*}

Having established the usability-relevance criteria, we developed three additional coding dimensions for the usability-relevant issues.
For RQ1, we characterize the nature of reported usability concerns using inductively generated themes and deductive mappings to Nielsen's usability heuristics.
For RQ2, we use affected-surface codes to identify where within the signing ecosystem each reported concern manifests.

\paragraph{Generating Inductive Usability Themes.}
Following the codebook-based thematic-analysis procedure described above~\cite{guest2011applied}, Analyst~A inductively grouped usability-relevant issues into themes based on similarities in the reported barriers and their implications for completing signing, verification, integration, or maintenance tasks.
A single issue could be assigned to multiple themes when the discussion exposed distinct usability concerns.
For example, an issue may begin as user confusion and later reveal a missing feature, in which case it is assigned to both a confusion-related theme and a missing-feature theme~\cite{notation_issue_1254}.
Analyst~A iteratively refined the themes and their definitions, while Analyst~B reviewed the themes after each revision stage and provided feedback on their interpretation and boundaries.
Analyst~A incorporated this feedback into subsequent revisions.


We further organized the inductive themes into broader \emph{theme families} that capture the underlying type of usability barrier represented in each issue.
For example, \emph{Operational Friction} groups concerns involving setup, configuration, integration, and workflow execution, while \emph{Cognitive Friction} groups concerns involving difficulty understanding tool behavior, guidance, or feedback.
Each theme maps to a predefined theme family in the codebook.

The complete theme and theme-family definitions are provided in ~\cref{tab:rq1_theme_definitions} and ~\cref{sec:appendix-methodology-detail}.

\paragraph{Mapping Nielsen's Usability Heuristics}
\label{para:method_nielsen}

We complemented the inductive themes with a deductive framework-mapping analysis based on Nielsen's usability heuristics.
Following framework-analysis approaches, which organize qualitative evidence using a predefined analytical framework~\cite{srivastava2009framework}, we treated Nielsen's ten heuristics as the a priori categories against which each usability-relevant issue was examined~\cite{Nielsen_1993}.
Although originally proposed for user-interface evaluation, the heuristics have also been applied to complex, domain-specific applications~\cite{kaplan2021_heuristics_complex_apps} and in prior usability research, including studies using software-repository data~\cite{Wibisono_Tjhin_2025,gonzalez2017examination,Sanei_Cheng_2024}.
This deductive mapping provides a second lens for RQ1, allowing concerns that emerge inductively from identity-based signing issues to also be organized using an established usability framework.

For each usability-relevant issue, we reviewed the full issue thread and identified the user-facing barrier evidenced by the initial report, subsequent discussion, and apparent resolution.
We then compared this evidence against the operational definition of each Nielsen heuristic in our codebook and assigned a heuristic when the reported barrier corresponded to the usability concern represented by that category.
Because an issue could exhibit more than one type of usability breakdown, multiple heuristic categories could be assigned when supported by distinct evidence in the discussion.
For example, unclear or non-actionable error feedback may map to N9 (\emph{Help Users Recognize, Diagnose, and Recover from Errors}), while insufficient or unclear usage guidance may map to N10 (\emph{Help and Documentation}).
The Nielsen mapping was conducted independently of the inductively generated themes rather than inferred from those theme labels.
\Cref{tab:nielsen_heuristics} summarizes the operational definitions used for this framework mapping; complete definitions and coding examples are provided in \cref{sec:appendix-methodology-detail}.

\begin{table*}[t]
\centering
\caption{Nielsen usability heuristics used for deductive coding.
Definitions are shortened here; complete operational definitions and coding examples are provided in \cref{sec:appendix-methodology-detail}.}
\label{tab:nielsen_heuristics}
\footnotesize
\setlength{\tabcolsep}{5pt}
\renewcommand{\arraystretch}{1.10}
\begin{tabularx}{\textwidth}{@{}p{0.1\textwidth} p{0.3\textwidth} X@{}}
\toprule
\textbf{Code} & \textbf{Heuristic} & \textbf{Operational Cue} \\
\midrule
N1 & Visibility of System Status & Clear and timely feedback about system state. \\
N2 & Match Between System and the Real World & Familiar language and concepts aligned with user expectations. \\
N3 & User Control and Freedom & Clear ways to exit, undo, or recover from unwanted states. \\
N4 & Consistency and Standards & Consistent terminology, behavior, and ecosystem conventions. \\
N5 & Error Prevention & Guardrails that help prevent avoidable mistakes. \\
N6 & Recognition Rather than Recall & Visible information and options that reduce memory burden. \\
N7 & Flexibility and Efficiency of Use & Support for efficient and adaptable workflows, including capabilities needed to complete or streamline legitimate tasks. \\
N8 & Aesthetic and Minimalist Design & Avoidance of irrelevant or distracting information. \\
N9 & Help Users Recognize, Diagnose, and Recover from Errors & Clear, specific, and actionable error and recovery guidance. \\
N10 & Help and Documentation & Discoverable, accurate, and useful help and documentation. \\
\bottomrule
\end{tabularx}
\end{table*}

\paragraph{Defining and Coding Affected Surfaces.}
For RQ2, we coded the \emph{affected surface} of each usability-relevant issue to identify where within the identity-based signing process the reported concern manifested.
As defined in \cref{sec:Background:conceptual_usability_identitySigning}, an affected surface denotes the component, interface, workflow, or integration boundary at which a usability concern becomes observable to the user.
The affected-surface categories were grounded in the architectural decomposition of identity-based signing systems summarized in \cref{tab:id_components} (and ~\cref{fig:bg_id_workflow}) and operationalized in the codebook.

For each issue, we reviewed the title, body, repository labels, associated comments, and apparent resolution to determine which surface or surfaces were substantively implicated by the reported barrier.
We coded the surface where the usability problem manifested rather than relying solely on the repository in which the issue was filed, because an issue filed in one repository (\eg Sigstore's Rekor) could concern another component or an interaction between components (\eg interaction between Rekor Logs and validating Identities from an OIDC Provider).
Affected-surface coding was multi-label when the discussion directly implicated more than one surface.
The complete affected-surface definitions and coding examples are provided in \cref{sec:appendix-methodology-detail}.

Together, the usability-relevance criteria, inductive themes, Nielsen heuristic mappings, and affected-surface definitions formed the finalized coding scheme used for LLM-assisted scaling and subsequent analysis.

\subsubsection{LLM-Assisted Coding At Scale}
\label{sec:method_llm_coding}

To scale coding beyond the human-annotated sample, we used an LLM as a co-annotator, following prior work on LLM-assisted qualitative coding and text annotation~\cite{Alizadeh_Kubli_Samei_Dehghani_Zahedivafa_Bermeo_Korobeynikova_Gilardi_2024,peng2025agentsperformcodeoptimization,wang2024human}.
The use of LLMs in qualitative analysis remains methodologically contested, particularly for reflexive forms of thematic analysis in which interpretation depends on researchers' situated judgment, subjectivity, and continued engagement with the data~\cite{vikan2026reflecting,alves2026llm}.
Our use differs from this more interpretive setting.
The LLM did not generate the analytical framework or independently construct themes; instead, it applied a human-developed codebook, predefined decision rules, and fixed annotation categories to additional issue threads.
This task is therefore closer to structured codebook-based and deductive coding, for which prior work has explored LLM assistance using predefined human-generated coding schemes~\cite{tai2024examination}.
Human analysts retained responsibility for developing and refining the coding scheme, defining theme boundaries, interpreting ambiguous cases, and validating the LLM's application of the finalized scheme.

The 100-issue development set was used to refine the annotation instructions and provide human-coded examples, while the separately held-out 80-issue set was reserved for independent evaluation of the finalized procedure.
Across the full 180-issue human-coded sample, 131 issues (72.8\%; approximate 95\% CI: 66.4\%--79.1\%, with finite-population correction) were adjudicated as usability-relevant under the operational definition in \cref{tab:method_usability_operationalization}.

\paragraph{Instruction Compilation From Human Coding.}
After completing the human coding-scheme development, we produced an instruction document and codebook that operationalized each annotation task.
The document specifies code definitions, decision rules, examples, and common edge cases for interpreting issue content, classifying usability relevance, assigning inductive themes and Nielsen heuristics, and identifying affected surfaces.

\paragraph{Iterative Calibration On The 180-Issue Set.}
We refined the annotation procedure using GPT~5.1 (API model) and the 100 human-coded development issues.
We supplied the full issue thread, including the title, repository labels, body, and comments, and instructed the model to proceed through three annotation stages: (1) issue understanding and affected-surface identification, (2) usability-relevance classification, and (3) inductive-theme and Nielsen-heuristic mapping.
Analyst~A reviewed the model outputs during development, documented recurring failure modes, and refined the instruction document to clarify ambiguous cases and reinforce decision rules~\cite{Khalid_Witmer_2025}.
We ran the scaled API classification with temperature set to 0 to reduce output variability and support auditability.
The 80 held-out issues were not used to refine the annotation instructions before the validation reported below.
To support reproducibility, the final classifier prompt, LLM output-processing scripts, and derived labels are included in the artifact repository described in \cref{sec:data_availability}.

\paragraph{Held-Out Validation.}
After finalizing the annotation procedure, we compared the LLM outputs against the human annotations for the separately held-out set of 80 issues.

\emph{Phase 1: Issue interpretation and affected surfaces.}
For issue understanding and affected-surface identification, we report percent agreement because this phase primarily evaluates whether the model interpreted the issue context and implicated surfaces consistently with the human annotations.
Agreement was 91.9\% for issue understanding and 66.7\% for affected-surface assignments.

\emph{Phase 2: Usability relevance.}
We compared human and LLM classifications of usability relevance.
Given the skew toward usability-relevant issues in the human-coded sample, we report Gwet's AC1~\cite{wongpakaran2013_comparison_kappa_ac1}, which is more stable than Cohen's $\kappa$ under skewed prevalence.
We obtained AC1 = 0.957 and Cohen's $\kappa = 0.856$.
The classifier achieved precision = 98.08\%, recall = 98.08\%, and F1 = 98.08\%, with specificity = 87.50\%.

\emph{Phase 3: Multi-label annotations.}
We evaluated agreement for the Nielsen heuristic, affected-surface, and inductive-theme assignments.
Percent agreement was 83.9\% for Nielsen heuristics, 75.8\% for affected surfaces, and 62.5\% for inductive themes.
For chance-corrected agreement, Gwet's AC1 was 0.84 for Nielsen heuristics, 0.86 for affected surfaces, and 0.91 for inductive themes.

After validation, we applied the finalized annotation procedure to the remaining issues in the corpus.
The resulting annotations form the basis of the analyses described next.


\subsection{Data Analysis}
\label{sec:analysis}

After full-corpus annotation, 2{,}965 of the 3{,}900 issue threads (76.0\%) were classified as usability-relevant.
We analyze these issues according to the coding dimensions described in \cref{sec:method_processing} to answer RQ1 and RQ2, and use their temporal distribution to answer RQ3.
We describe the analysis for each research question below.

\subsubsection{RQ1: Characterizing Reported Usability Concerns}


RQ1 asks what usability problems are reported for identity-based software signing tools.
To answer RQ1, we analyze two complementary representations of the reported usability concerns.
First, we summarize the frequencies and distributions of the inductively generated usability themes and their broader theme families to characterize recurring problem types that emerge from the identity-based signing context.
Second, we summarize the Nielsen heuristic annotations to characterize the same concerns using an established usability framework.
For both representations, we examine the overall distributions and compare their prevalence across the individual signing ecosystems to identify recurring and tool-specific patterns.
Because issues may receive multiple theme and heuristic labels, the reported percentages need not sum to 100\%. Also, because theme-family membership is multi-label, we use issue-level logistic GEE, clustering observations by issue, to test whether family-prevalence patterns differ across repositories~\cite{liang1986longitudinal}; full analysis outputs are included in our artifact.

\subsubsection{RQ2: Identifying Affected Surfaces}

RQ2 asks which affected surfaces are most frequently implicated in reported usability problems.
To answer RQ2, we summarize the frequencies and distributions of affected-surface annotations across the corpus and compare these distributions across signing ecosystems.
This analysis identifies the components, interfaces, workflows, and integration boundaries most frequently associated with reported usability concerns.
Because an issue may implicate more than one surface, affected-surface categories are non-mutually exclusive and their percentages need not sum to 100\%.

\subsubsection{RQ3: Longitudinal Trends in Reported Usability Concerns}
RQ3 asks how user-reported usability concerns change over time in identity-based software signing tools.
Guided by the \emph{Tool Usability Lifecycle} introduced in \cref{sec:Background:UsabilityTheory}, we conduct an exploratory analysis of whether reported concerns increase, remain relatively stable, or decline over the observation period.
We examine these trajectories for the aggregated corpus and compare them across tools, inductive usability themes, and affected surfaces.
We interpret issue-count trends as observable public signals of reported usability friction rather than direct measurements of total usability burden, user experience, adoption, or tool maturity.

We construct monthly time series of usability-issue counts over the study window.
Because the outcome is a nonnegative integer count per month, we model temporal trends using Poisson regression.
Poisson regression directly models count outcomes, ensures nonnegative fitted values, and allows the effect of time to be interpreted multiplicatively through the log link~\cite{cameron_trivedi_2013}.
For each modeled series, we estimate:

\begin{equation}
\label{eq:poisson}
\log(\mathbb{E}[Y_t]) = \beta_0 + \beta_1 t,
\end{equation}

where $Y_t$ is the number of user-reported usability issues in month $t$, $\beta_0$ is the intercept, and $\beta_1$ represents the estimated log-linear change in the expected issue count per month.
We report $\exp(\beta_1)$ as the corresponding monthly rate ratio.
A negative $\beta_1$ indicates a declining fitted trend, a positive value indicates an increasing fitted trend, and estimates near zero indicate little net change over the observation window.
This specification therefore characterizes the net temporal direction within our fixed observation window rather than fitting a complete lifecycle trajectory.

We first apply the model to the aggregated usability-issue series to characterize the overall temporal pattern.
We then apply the same model separately across tools to examine whether their trajectories differ.
Finally, we repeat the analysis for inductive usability themes and affected surfaces to examine whether different concern types and parts of the signing ecosystem exhibit different temporal patterns.

Because the \emph{Tool Usability Lifecycle} allows for multiple plausible temporal patterns, including increasing, stable, and declining reported concerns, we treat these trend analyses as exploratory.
This is particularly important for the disaggregated theme- and affected-surface analyses, which involve many related trend estimates.
We therefore focus on the direction, magnitude, uncertainty, and consistency of the estimated trends rather than on isolated statistically significant results.
For the aggregate theme- and affected-surface models, we retain the Poisson specification for estimating temporal slopes and fitted counts, but base inference primarily on HC0 heteroskedasticity-robust standard errors to reduce sensitivity to violations of the Poisson mean--variance assumption.
Negative-binomial models provide an additional sensitivity check.
We report full Poisson diagnostics and robustness analyses in Appendix~\ref{sec:appendix-methodology-detail-poisson-assumption-check}.


\section{Results}
\label{sec:results}


In this section, we present our findings by research question.
~\cref{sec:results:RQ1} characterizes the usability problems reported across the studied identity-based signing ecosystems (RQ1).
~\cref{sec:results_rq2} identifies the affected surfaces where these problems most frequently manifest (RQ2).
~\cref{sec:results_rq3} examines how reported usability concerns change over the observation period (RQ3).

Across the full corpus, 2{,}965 of 3{,}900 issue threads were classified as usability-relevant.
Usability-related concerns therefore constitute a substantial share of the issues in our filtered corpus under the workflow-level operationalization of usability described in \cref{sec:method_processing}.
This proportion should not be interpreted as the prevalence of usability problems among all users or tool interactions because our denominator consists of publicly reported GitHub issue threads rather than usage events.
As a descriptive check, the full-corpus proportion is similar to the 72.8\% observed in the human-coded sample and falls within its approximate 95\% confidence interval (66.4\%--79.1\%).

\subsection{RQ1: Reported Usability Problems}
\label{sec:results:RQ1}


\textbf{RQ1} asks what usability problems are reported for identity-based software signing tools.
We answer this through two complementary views.
In~\cref{sec:results:RQ1:inductive}, we characterize the reported concerns using the inductively generated usability themes and theme families.
In~\cref{sec:results:RQ1:nielsen}, we examine the corresponding Nielsen heuristic mappings as a deductive complement.

\begin{tcolorbox}[findingsbox]
\textbf{Finding 1:}
Overall, RQ1 shows that reported usability problems are broad but patterned.
Operational workflow friction, unmet functional expectations, and information clarity concerns occur at similar rates, while reliability breakdowns are less frequent.
These trends recur across all tools.
\end{tcolorbox}

\subsubsection{Inductively Generated Usability Themes}
\label{sec:results:RQ1:inductive}

Our inductive analysis identified eleven recurring usability themes, which we organize into four broader \emph{theme families}: \emph{Functional Gap}, \emph{Operational Friction}, \emph{Cognitive Friction}, and \emph{Functional Reliability}.
\Cref{tab:rq1_theme_definitions} summarizes the themes, their definitions, and the broader families to which they belong; complete coding definitions and examples are provided in Appendix~\cref{tab:phase3-theme-codebook}.
Because issues could receive multiple theme labels, the counts below represent assignments rather than mutually exclusive issue categories.
An issue may receive multiple themes within the same predefined family, but that family is counted only once for the issue; consequently, the corpus contains 5{,}924 theme assignments and 5{,}458 theme-family assignments.
\Cref{tab:l1-theme-grouped-percent-by-tool} reports the per-tool distributions.

\begin{table*}[t]
\centering
\caption{Inductively generated usability themes and their higher-level theme families used in RQ1.
Theme families group related usability themes at a higher level of abstraction.
Theme definitions are shortened from the complete coding definitions reported in \cref{sec:appendix-methodology-detail}.}
\label{tab:rq1_theme_definitions}

\footnotesize
\setlength{\tabcolsep}{5pt}
\renewcommand{\arraystretch}{1.30}

\begin{tabularx}{\textwidth}{
@{}
>{\raggedright\arraybackslash}p{0.22\textwidth}
>{\raggedright\arraybackslash}p{0.27\textwidth}
>{\raggedright\arraybackslash}X
@{}
}
\toprule
\textbf{Theme Family} & \textbf{Theme} & \textbf{Theme Definition} \\
\midrule

\textbf{Functional Gap}\newline
\emph{Captures cases where a capability needed for an intended workflow is absent or insufficient.}
&
Missing Feature/Enhancement Request
&
A needed capability does not exist, or an existing capability requires enhancement to support the reported task.
\\

\midrule

\multirow[c]{5}{0.22\textwidth}{%
\parbox{0.21\textwidth}{\raggedright
\textbf{Operational Friction}\\[3pt]
\emph{Captures cases where a capability exists, but setup, authentication, integration, or workflow mechanics make the task difficult to execute.}
}}
&
Configuration Friction
&
Users have difficulty configuring the tool because of complex, rigid, unclear, or insufficiently documented requirements.
\\

&
Authentication Friction
&
Authentication or authorization creates barriers through identity setup, token handling, credential flows, timeouts, or related requirements.
\\

&
Build/CI/Installation/\allowbreak Distribution/\allowbreak Release Issues
&
Users encounter barriers when installing, building, distributing, releasing, or operating the tool in local or CI/CD environments.
\\

&
Integration Failure/Issues
&
The tool is difficult to integrate or interoperate with another component, platform, or external system required by the workflow.
\\

&
Tedious Workflows
&
Completing an intended task requires unnecessary, repetitive, manual, or otherwise burdensome steps.
\\

\midrule

\multirow[c]{2}{0.22\textwidth}{%
\parbox{0.21\textwidth}{\raggedright
\textbf{Cognitive Friction}\\[3pt]
\emph{Captures cases where users struggle to understand tool behavior, guidance, or feedback.}
}}
&
User Confusion/Unclear Documentation
&
Users have difficulty understanding tool behavior, expected usage, or instructions because guidance is unclear, incomplete, outdated, or unavailable.
\\

&
Notification/Logging Issues
&
Errors, logs, notifications, or other system feedback are unclear, incomplete, noisy, or insufficiently actionable.
\\

\midrule

\multirow[c]{3}{0.22\textwidth}{%
\parbox{0.21\textwidth}{\raggedright
\textbf{Functional Reliability}\\[3pt]
\emph{Captures cases where a capability exists, but its behavior, performance, or security properties undermine dependable use.}
}}
&
Unexpected Behavior
&
The tool behaves differently from what users expect or from what the documented workflow indicates.
\\

&
Performance Issues
&
The tool performs its intended function, but latency, resource use, rate limits, or related performance characteristics impede practical use.
\\

&
Security Concerns
&
The reported behavior raises concerns about vulnerabilities, unsafe defaults, information exposure, trust guarantees, or other security properties.
\\

\bottomrule
\end{tabularx}
\end{table*}

\begin{table*}[t]
\centering
\caption{Usability theme distribution by tool, normalized to show the percentage of theme assignments per tool. 
Notation:
  \textit{Config}---Config. friction;
  \textit{Auth}---Authentication friction;
  \textit{Build/Rel}---Build/CI/install/distribution issues;
  \textit{Integration}---Integration;
  \textit{Workflow}---Tedious Workflows;
  \textit{Docs}---User confusion/unclear docs;
  \textit{Notif/Logs}---Notification/Logging;
  \textit{Unexp Beh}---Unexpected behavior;
  \textit{Performance}---Performance;
  \textit{Security}---Security;
  \textit{Missing}---Missing feature/enhancement.
  }
\label{tab:l1-theme-grouped-percent-by-tool}
\scriptsize
\small
\setlength{\tabcolsep}{2.5pt}
\renewcommand{\arraystretch}{1.15}
\begin{tabular}{c|ccccc|cc|ccc|c}
\toprule
\textbf{Tool} 
& \multicolumn{5}{c|}{\textbf{Operational Friction}}
& \multicolumn{2}{c|}{\textbf{Cognitive Friction}}
& \multicolumn{3}{c|}{\textbf{Functional Reliability}}
& \multicolumn{1}{c}{\textbf{Functional Gap}} \\
\midrule
& \rotatebox{90}{\textit{Config}}
& \rotatebox{90}{\textit{Auth}}
& \rotatebox{90}{\textit{Build/Rel}}
& \rotatebox{90}{\textit{Integration}}
& \rotatebox{90}{\textit{Workflow}}
& \rotatebox{90}{\textit{Docs}}
& \rotatebox{90}{\textit{Notif/Logs}}
& \rotatebox{90}{\textit{Unexp Beh.}}
& \rotatebox{90}{\textit{Performance}}
& \rotatebox{90}{\textit{Security}}
& \rotatebox{90}{\textit{Missing}} \\
\midrule
Sigstore-Cosign      &  4.4 &  2.6 &  4.7 & 13.9 &  6.5 & 18.5 &  4.4 &  9.9 & 1.0 &  4.8 & 29.4 \\
Sigstore-Fulcio      &  3.9 &  1.8 &  4.3 & 13.6 &  2.9 & 11.4 &  4.3 &  6.8 & 1.1 & 11.1 & 38.9 \\
Sigstore-Rekor       &  2.2 &  0.0 &  4.4 & 10.2 &  4.0 & 16.8 &  5.0 & 10.2 & 3.7 &  6.5 & 37.0 \\
\midrule
Keyfactor-SignServer & 11.6 &  9.3 & 11.6 &  9.3 &  2.3 & 30.2 &  7.0 &  4.6 & 2.3 &  0.0 & 11.6 \\
Keyfactor-EJBCA      & 14.7 &  3.9 & 12.4 & 10.5 &  1.3 & 27.4 &  8.5 & 11.1 & 0.6 &  2.0 &  7.5 \\
\midrule
Notary v2-Notation   &  4.2 &  1.5 &  5.0 &  7.4 & 10.1 &  8.8 & 12.6 &  6.3 & 1.7 &  3.6 & 38.7 \\
OpenPubKey           &  1.3 &  3.3 &  5.3 &  7.3 &  2.7 & 10.0 &  4.0 &  6.0 & 0.7 & 10.7 & 48.7 \\
HashiCorp-Vault      &  7.2 &  4.4 &  3.6 & 10.6 &  7.9 & 16.4 & 11.1 & 15.1 & 1.9 &  3.0 & 18.9 \\
\bottomrule
\end{tabular}
\end{table*}

At the issue level, theme-family prevalence patterns differ significantly across repositories
(Wald $\chi^2(21)=378.11$, $p<.001$).

\myparagraph{Functional Gap — users report missing affordances needed to complete signing workflows.}
\label{sec:results:RQ1:gap}

Functional Gap accounts for 1{,}481 theme-family assignments (27.1\%).
Its constituent theme, \emph{Missing Feature/Enhancement Request}, is also the most frequent individual theme in the corpus ($N=1{,}481$; 25.0\% of all theme assignments).
These issues describe cases in which a capability or affordance needed to support an intended signing, verification, integration, or maintenance workflow is absent or insufficient.
For example, \href{https://github.com/sigstore/cosign/issues/4119}{Cosign~\#4119} requests URL support for a verification bundle to avoid a manual download step in automated workflows, while \href{https://github.com/Keyfactor/ejbca-ce/issues/943}{EJBCA~\#943} requests CLI support for specifying an alternative signature algorithm when creating a hybrid CA.
In identity-based signing, such gaps are consequential because signing and verification are commonly embedded in automated release and CI/CD workflows; a missing capability can therefore force manual workarounds or prevent the intended workflow from being completed efficiently.

\emph{Missing Feature/Enhancement Request} accounts for a particularly large share of theme assignments in OpenPubKey (48.7\%), Fulcio (38.9\%), Notation (38.7\%), and Cosign (29.4\%).
Its prevalence also varies more across tools than that of the other individual themes (Cram\'er's $V=0.215$), indicating that reported capability gaps are distributed unevenly across the studied tools.

\myparagraph{Operational Friction — users struggle at configuration, authentication, build/release, and integration boundaries.}

Operational Friction is the largest theme family, accounting for 1{,}534 theme-family assignments (28.1\%).
Its most frequent themes are \emph{Integration Failure/Issues} ($N=659$; 11.1\%), \emph{Tedious Workflows} ($N=401$; 6.8\%), and \emph{Configuration Friction} ($N=364$; 6.1\%), followed by \emph{Build/CI/Installation/Distribution/Release Issues} ($N=272$; 4.6\%) and \emph{Authentication Friction} ($N=199$; 3.4\%).
Together, these themes show that a substantial share of reported concerns arise when users attempt to configure, authenticate, integrate, and operate signing tools within existing development and release environments.
Representative reports include failures when interoperating with adjacent systems (\emph{Integration Failure/Issues}; \href{https://github.com/notaryproject/notation/issues/1254}{Notation~\#1254}), verification blocked by configuration requirements (\emph{Configuration Friction}; \href{https://github.com/notaryproject/notation/issues/430}{Notation~\#430}), and authentication flows that are difficult to use in non-interactive CI environments (\emph{Authentication Friction}; \href{https://github.com/sigstore/cosign/issues/1108}{Cosign~\#1108}).

Identity-based signing reduces long-lived key-management burden, but it shifts more of the workflow onto identity providers, policy configuration, and automated infrastructure.
When these surrounding systems are difficult to configure or integrate, signing and verification become harder to deploy reliably in practice.

\myparagraph{Cognitive Friction: users lack clear guidance, feedback, and recoverability.}
\label{sec:results:RQ1:cognitive}

Cognitive Friction accounts for 1{,}428 theme-family assignments (26.2\%).
\emph{User Confusion/Unclear Documentation} is the second-most frequent individual theme overall ($N=982$; 16.6\%), while \emph{Notification/Logging Issues} accounts for 512 assignments (8.6\%).
These reports concern difficulty understanding expected tool behavior, usage guidance, system feedback, or recovery actions.
For example, verification documentation that omits required certificate-identity information (\emph{User Confusion/Unclear Documentation}; \href{https://github.com/sigstore/cosign/issues/3671}{Cosign~\#3671}) and verification feedback that obscures the underlying trust-policy mismatch (\emph{Notification/Logging Issues}; \href{https://github.com/notaryproject/notation/issues/625}{Notation~\#625}) illustrate the two main forms of Cognitive Friction.

This type of friction matters for identity-based signing because users must correctly interpret several pieces of security-relevant information, including identity claims, certificate and issuer information, trust-policy requirements, and verification output.
When guidance or feedback is unclear, users may struggle to distinguish a genuine verification failure from a configuration or workflow problem, determine which component caused the failure, or identify the action needed to recover.
This can make it harder to configure trust policies correctly and to make informed decisions about whether a signing or verification result should be accepted.

\myparagraph{Functional Reliability: users encounter unexpected behavior, performance degradation, and security-related breakdowns.}
\label{sec:results:RQ1:reliability}

Functional Reliability is the smallest of the four theme families, accounting for 1{,}015 theme-family assignments (18.6\%).
Its dominant theme is \emph{Unexpected Behavior} ($N=709$; 12.0\%), followed by \emph{Security Concerns} ($N=248$; 4.2\%) and \emph{Performance Issues} ($N=97$; 1.6\%).
Representative reports include behavior inconsistent with documented signing support (\emph{Unexpected Behavior}; \href{https://github.com/sigstore/fulcio/issues/1858}{Fulcio~\#1858}), uncertainty about signing-algorithm behavior in KMS integrations (\emph{Security Concerns}; \href{https://github.com/sigstore/cosign/issues/3807}{Cosign~\#3807}), and certificate-enrollment performance degradation (\emph{Performance Issues}; \href{https://github.com/Keyfactor/ejbca-ce/issues/211}{EJBCA~\#211}).

Although less frequent than the other families, reliability concerns matter for identity-based signing because users depend on signing and verification workflows to produce predictable security-relevant outcomes.
Unexpected behavior can make it unclear whether a signature was created or verified as intended, while performance problems can make signing or verification impractical in automated or high-volume workflows.
Security-related uncertainty can make it difficult for users to determine whether the system is applying the expected algorithms, policies, or trust guarantees.
These concerns are especially difficult to diagnose in identity-based signing because failures may span several interacting components, including the signing tool, identity provider, credential issuer, policy configuration, and verification infrastructure.

Vault and EJBCA show the largest Functional Reliability shares (19.5\% and 17.4\%, respectively).
Overall, reliability-related concerns are less frequent than Functional Gap, Operational Friction, and Cognitive Friction in the observed issue corpus.

\subsubsection{Nielsen Heuristic Mapping: A Deductive Complement}
\label{sec:results:RQ1:nielsen}

\begin{table}[t]
\centering
\caption{
Distribution of Nielsen usability heuristic assignments (N1--N10, cf.~\cref{sec:method_human_annotation}) by tool (percent of heuristic assignments per tool). N7 (\emph{Flexibility and Efficiency of Use}) and N10 (\emph{Help and Documentation}) are the two most common heuristics across tools. Bold values indicate the top three heuristics within each tool/component row. The \emph{Average} row reports the unweighted mean of the eight tool/component percentages.}
\label{tab:nielsen-theme-percent-by-tool}
\small
\setlength{\tabcolsep}{2.5pt}
\renewcommand{\arraystretch}{1.1}
\begin{tabular*}{\columnwidth}{@{\extracolsep{\fill}}lrrrrrrrrrr}
\toprule
\textbf{Tool} 
& \textbf{N1}
& \textbf{N2}
& \textbf{N3}
& \textbf{N4}
& \textbf{N5}
& \textbf{N6}
& \textbf{N7}
& \textbf{N8}
& \textbf{N9}
& \textbf{N10} \\
\midrule
Cosign      &  3.5 &  6.6 &  2.6 & 12.3 &  9.2 &  2.6 & \textbf{27.4} &  1.8 & \textbf{12.9} & \textbf{21.1} \\
Fulcio     &  4.1 &  5.1 &  0.7 & \textbf{12.0} & \textbf{12.0} &  2.4 & \textbf{33.9} &  2.4 &  6.5 & \textbf{20.9} \\ 
Rekor       &  4.6 &  7.2 &  0.3 & \textbf{12.7} &  8.9 &  2.0 & \textbf{32.3} &  1.2 & 11.0 & \textbf{19.9} \\
\midrule
SignServer  &  2.2 &  0.0 &  0.0 &  6.7 & 11.1 &  0.0 & \textbf{15.6} &  2.2 & \textbf{24.4} & \textbf{37.8} \\
EJBCA       &  4.8 &  4.8 &  1.1 & 10.5 & \textbf{14.0} &  0.3 & 12.5 &  0.3 & \textbf{21.1} & \textbf{30.5} \\
\midrule
Notation    &  8.5 &  5.7 &  1.6 & \textbf{12.7} &  8.5 &  4.8 & \textbf{26.9} &  2.8 & 10.3 & \textbf{18.2} \\
OpenPubKey  &  0.6 &  4.2 &  3.0 &  9.7 & \textbf{17.6} &  1.8 & \textbf{38.8} &  3.6 &  6.1 & \textbf{14.5} \\
Vault       &  7.3 &  6.2 &  2.5 & 12.5 & 11.4 &  1.9 & \textbf{20.3} &  1.3 & \textbf{15.9} & \textbf{20.7} \\
\midrule
\textit{Average} & 4.4 & 5.0 & 1.5 & 11.2 & 11.3 & 2.0 & \textbf{26.5} & 2.0 & \textbf{12.7} & \textbf{23.3} \\
\bottomrule
\end{tabular*}
\begin{flushleft}
\scriptsize
\textit{Note.} N1=Visibility of System Status; N2=Match Between System and the Real World; N3=User Control and Freedom; N4=Consistency and Standards; N5=Error Prevention; N6=Recognition Rather than Recall; N7=Flexibility and Efficiency of Use; N8=Aesthetic and Minimalist Design; N9=Help Users Recognize, Diagnose, and Recover from Errors; N10=Help and Documentation.
\end{flushleft}
\end{table}


The Nielsen heuristic analysis provides a complementary view of the inductive findings using an established usability framework.
\Cref{tab:nielsen-theme-percent-by-tool} shows the overall and per-tool distributions.

\emph{N7~---~Flexibility and Efficiency of Use} (23.8\%) and
\emph{N10~---~Help and Documentation} (21.0\%) rank first and
second.
These broadly correspond to the workflow-efficiency and missing-capability concerns represented by Operational Friction and Functional Gap, and the documentation-related concerns represented by Cognitive Friction.
\emph{N9~---~Help Users Recognize, Diagnose, and Recover from Errors} (14.1\%) ranks third, reflecting difficulties interpreting failures and identifying appropriate recovery actions.
\emph{N4~---~Consistency and Standards} (12.2\%) and
\emph{N5~---~Error Prevention} (10.8\%) round out the top five, highlighting additional concerns related to consistency and preventing errors in signing and verification workflows.
These results broadly align with the inductive analysis.
N7 reflects many of the workflow-efficiency and missing-capability concerns represented by Operational Friction and Functional Gap, while N10 and N9 reflect documentation, feedback, and recovery concerns represented by Cognitive Friction.

The two analyses therefore provide complementary perspectives on the same corpus.
The inductive themes characterize recurring usability concerns specific to identity-based signing, while the Nielsen mapping relates those concerns to established usability principles.
Their broad alignment suggests that many of the domain-specific barriers observed in identity-based signing arise through familiar usability mechanisms, such as inefficient workflows, inadequate guidance, unclear feedback, and difficult error recovery.
What makes these barriers distinctive in this setting is that they occur within security-sensitive, multi-component workflows involving identity, policy, certificates, and verification decisions.
We return to this interaction between general usability principles and signing-specific complexity in \cref{sec:discussion}.
\subsection{RQ2: Affected Surfaces Implicated
in Usability Reports}
\label{sec:results_rq2}

\textbf{RQ2} asks where reported usability problems appear within identity-based signing tools.
It complements \emph{RQ1}: while \emph{RQ1} characterizes the nature of the reported problems, \emph{RQ2} identifies the affected surfaces where they manifest.
Following the terminology introduced in \cref{sec:Background:conceptual_usability_identitySigning}, an \emph{affected surface} denotes the component, interface, workflow, or integration boundary at which a usability concern becomes observable to the user.

Affected-surface annotations are multi-label, yielding 5{,}722 surface assignments across the 2{,}965 usability-relevant issues.
\Cref{tab:rq2-components-top} reports the overall distribution, while \cref{tab:rq2-top3-components-by-tool} summarizes the leading surfaces for each repository.

\begin{tcolorbox}[findingsbox]
\textbf{Finding 2:}
\rev{RQ2 shows that reported usability concerns are concentrated in
(1) CLI tooling and API interaction surfaces,
(2) policy and configuration boundaries, and
(3) signing and verification workflow surfaces.
These results identify the surfaces and boundaries where usability friction most often appears in identity-based signing tools.}
\end{tcolorbox}

\begin{table}[t]
\centering
\caption{Top 10 affected surfaces across all studied repositories.
Counts denote affected-surface assignments, and shares represent the percentage of all affected-surface assignments ($N=5{,}722$).
The two omitted categories are \textit{General/Unspecified Component} 
(coded as \textit{Core}; $N=173$) and \textit{Release Pipeline} 
($N=117$), which together account for 5.1\% of all affected-surface assignments.
Because affected-surface coding is multi-label, a single usability-relevant issue may contribute more than one assignment.
Affected-surface categories are defined across the corpus and are not necessarily represented in every studied ecosystem.
\textit{Key Management Core / Secrets Backend} primarily reflects Vault-specific secrets-management infrastructure and is retained in the ranking for completeness.
Labels marked with $^{\dagger}$ denote components or interfaces, while labels marked with $^{\ast}$ denote end-to-end signing workflows.}
\label{tab:rq2-components-top}
\small
\setlength{\tabcolsep}{5pt}
\renewcommand{\arraystretch}{1.12}
\begin{tabular}{@{}lrr@{}}
\toprule
\textbf{Affected Surface} & \textbf{Count} & \textbf{Share (\%)} \\
\midrule
CLI Tooling$^{\dagger}$                  & 1{,}110 & 19.4 \\
Policy/Configuration$^{\dagger}$         &   847   & 14.8 \\
API$^{\dagger}$                          &   654   & 11.4 \\
Key Management Core / Secrets Backend$^{\dagger}$ & 563  &  9.8 \\
Authentication/Authorization$^{\dagger}$ &   495   &  8.7 \\
Signing Workflow$^{\ast}$             &   407   &  7.1 \\
Verification Workflow$^{\ast}$        &   390   &  6.8 \\
Web Client$^{\dagger}$                   &   372   &  6.5 \\
Build/CI/Installation$^{\dagger}$        &   340   &  5.9 \\
Notification/Logging$^{\dagger}$         &   254   &  4.4 \\
\bottomrule
\end{tabular}
\end{table}

\subsubsection{CLI and API Surfaces: Primary Developer-Facing Interaction Layer.}
\label{para:results:RQ2:cli-api}

The most frequently implicated affected surfaces are
\emph{CLI Tooling} ($N=1{,}110$; 19.4\%) and
\emph{API} ($N=654$; 11.4\%), which together account for 30.8\% of all surface assignments.
These surfaces expose signing and verification functionality directly to users and automated workflows.
Reported concerns include confusing or unsuitable CLI output, inconsistent API behavior, and interaction patterns that complicate automation and diagnosis.
At these surfaces, usability problems translate abstract trust concepts into concrete workflow friction: confusing flags, inconsistent request/response semantics, weak defaults for non-interactive use, or limited feedback can all make automation difficult to build and diagnose.
For example, scripted verification can require cleaner CLI output than the tool exposes
(\href{https://github.com/sigstore/cosign/issues/2832}{Cosign~\#2832}), while an API workflow can fail to provide data needed by infrastructure-as-code automation after a successful update
(\href{https://github.com/hashicorp/vault/issues/29150}{Vault~\#29150}).

Notation and Cosign show the largest within-tool concentrations of CLI-related concerns (36.2\% and 30.3\%, respectively).
By contrast, Rekor shows the largest concentration of API-related concerns (39.1\%), consistent with its programmatic interface to the transparency log.


\subsubsection{Policy and Configuration Surfaces: The Trust Specification Layer}
\label{sec:results:RQ2:policy}

\emph{Policy/Configuration} is the second-most frequently implicated affected surface ($N=847$; 14.8\%) and the leading surface for Vault (18.4\%), EJBCA (25.6\%), and Fulcio (22.2\%).
These reports concern difficulties specifying and maintaining the configuration and trust requirements that govern signing and verification.
For example, users request templates, defaults, and CLI support for trust-policy configuration rather than manually editing policy files
(\href{https://github.com/notaryproject/notation/issues/398}{Notation~\#398}).

Policy and configuration concerns are particularly important in identity-based signing because trust decisions depend on correctly coordinating identity, issuer, certificate, and verification-policy information.
When this configuration is incorrect or unclear, users may have difficulty determining why signing or verification does not behave according to the intended trust policy.

\subsubsection{Authentication and Identity Surfaces: The Trust Boundary Between Development Tooling and Identity Infrastructure.}
\label{sec:results:RQ2:auth}

\emph{Authentication/Authorization} accounts for $N=495$ surface assignments (8.7\%).
These reports concern difficulties with OIDC flows, tokens, issuer configuration, permissions, RBAC, and related authorization behavior.
For example, OpenPubKey~\#294 illustrates authentication friction caused by strict token-validity checks: authentication failed because of a three-minute client clock skew, requiring the user to manually correct the system clock before proceeding
(\href{https://github.com/openpubkey/openpubkey/issues/294}{OpenPubKey~\#294}).
Authentication concerns are most prominent in OpenPubKey (23.4\%) and Fulcio (15.4\%), where identity (OIDC-based authentication) plays a central role in the workflow.

These concerns are particularly important in identity-based signing because signing depends on coordination between the signing tool and external identity infrastructure.
When that coordination fails, users may have difficulty obtaining the credentials needed to sign or diagnosing why authentication was rejected.

\subsubsection{Signing and Verification Workflow Surfaces: Adoption-Critical Operations.}
\label{sec:results:RQ2:workflows}

\emph{Signing Workflow} and \emph{Verification Workflow} account for $N=407$ (7.1\%) and $N=390$ (6.8\%) surface assignments, respectively, or $N=797$ (13.9\%) combined.
These reports concern difficulties completing the end-to-end tasks of producing and verifying signed artifacts.
For example, some private-PKI signing workflows require timestamp responses to be manually attached alongside signatures (\href{https://github.com/sigstore/cosign/issues/2979}{Cosign~\#2979}).
Notation~\#625 illustrates verification friction caused by unclear error feedback: verification fails without identifying which trust-policy condition was violated, making it difficult for users to diagnose and correct the problem (\href{https://github.com/notaryproject/notation/issues/625}{Notation~\#625}).
Signing- and verification-workflow concerns are most prominent in Cosign and Notation, with verification accounting for 17.5\% and 12.4\% of their assignments and signing accounting for 17.5\% and 15.6\%, respectively.

These surfaces are particularly important because they represent the tasks through which users create signatures and determine whether signed artifacts satisfy their trust requirements.
Unclear or incomplete workflow behavior can therefore make it difficult to complete signing or understand why verification succeeds or fails.

\subsubsection{Remaining Surfaces: Supporting Signing Infrastructure}
\label{sec:results:RQ2:remaining}

Reported usability concerns also appear across supporting infrastructure used to deploy, operate, and diagnose signing systems.
\emph{Key Management Core / Secrets Backend} accounts for $N=563$ assignments (9.8\%) and is concentrated primarily in Vault, while \emph{Web Client} accounts for $N=372$ (6.5\%) and is concentrated in EJBCA and SignServer.
\emph{Build/CI/Installation} accounts for $N=340$ assignments (5.9\%), \emph{Notification/Logging} for $N=254$ (4.4\%), and \emph{Release Pipeline} for $N=117$ (2.0\%).
Representative reports include Vault~\#31601, where write-only secret editing available only through CLI workflows creates additional interaction steps for users
(\href{https://github.com/hashicorp/vault/issues/31601}{Vault~\#31601});
EJBCA~\#932, where an administrative page appears usable without authentication but then fails without clearly communicating the authentication requirement
(\href{https://github.com/Keyfactor/ejbca-ce/issues/932}{EJBCA~\#932});
Vault~\#29054, where missing packages prevent users from completing Docker-based installation workflows
(\href{https://github.com/hashicorp/vault/issues/29054}{Vault~\#29054}); and
Cosign~\#4438, where an OIDC failure is surfaced as a low-level parsing error rather than feedback that helps the user identify and correct the authentication problem
(\href{https://github.com/sigstore/cosign/issues/4438}{Cosign~\#4438}).

Together, these surfaces show that usability concerns extend beyond the signing and verification interfaces to the surrounding infrastructure needed to deploy, automate, and troubleshoot identity-based signing.

\subsection{RQ3: How do reported usability problems change over time?}
\label{sec:results_rq3}

We answer RQ3 in three steps.
First, we examine temporal trends in reported usability problems across the studied tools (\cref{sec:results:RQ3:tool_component}).
Second, we examine theme- and heuristic-level trends to identify which types of reported usability concerns change within each tool (\cref{sec:results:RQ3:tool_theme}).
Third, we pool issues across tools to examine overall trends by usability theme and affected surface (\cref{sec:results:RQ3:pooled}).

\begin{tcolorbox}[findingsbox]
\textbf{Finding 3:}
Reported usability concerns follow different trajectories across identity-based signing tools and concern types.
Monthly issue counts decline significantly for Cosign, Fulcio, Rekor, Vault, and Notation, increase for OpenPubKey, and show no significant overall change for EJBCA or SignServer.
Theme- and affected-surface analyses show further variation within these overall trajectories, indicating that different usability concerns change at different rates over the observation period.
\end{tcolorbox}

\subsubsection{Temporal Trends Across Tools}
\label{sec:results:RQ3:tool_component}

At the repository (tool-component) level (\cref{tab:poisson-overall-trends}), five of the eight repositories show significant declines in monthly usability-issue counts.
All three Sigstore components decline significantly:
Cosign ($\beta_1=-0.0382$, $p<0.001$),
Fulcio ($\beta_1=-0.0479$, $p<0.001$), and
Rekor ($\beta_1=-0.0505$, $p<0.001$).
Notation ($\beta_1=-0.0322$, $p<0.001$) and Vault ($\beta_1=-0.0250$, $p<0.001$) also show significant declines.

In contrast, OpenPubKey shows a significant increase in reported usability issues over the observation period
($\beta_1=0.0404$, $p<0.001$).
Neither Keyfactor repository shows a significant overall trend:
EJBCA ($\beta_1=-0.0001$, $p=0.991$) remains nearly flat, while SignServer has a small positive but non-significant slope
($\beta_1=0.0126$, $p=0.408$).

\begin{table}[t]
\centering
\caption{
Classical Poisson time-trend estimates for monthly counts of usability-related issues (all themes combined).
Negative $\beta_1$ indicates decreasing counts over time.
Trends significant under classical Poisson inference ($p<0.05$) are bolded.
Robust-standard-error checks and negative-binomial sensitivity models are reported in Appendix~\cref{sec:appendix-methodology-detail-poisson-assumption-check}.
}
\label{tab:poisson-overall-trends}
\scriptsize
\small
\footnotesize
\setlength{\tabcolsep}{4.5pt}
\renewcommand{\arraystretch}{1.12}
\begin{tabular}{@{}l c c c c r@{}}
\toprule
\textbf{Repository} & \textbf{$\beta_1$} & \textbf{RR/Month} & \textbf{\%/Month} & \textbf{$p$} & \textbf{Total} \\
\midrule
sigstore/cosign         & \textbf{-0.0382} & \textbf{0.963} & \textbf{-3.75} & \textbf{$2.43{\times}10^{-40}$} & \textbf{716} \\
sigstore/fulcio         & \textbf{-0.0479} & \textbf{0.953} & \textbf{-4.67} & \textbf{$4.14{\times}10^{-13}$} & \textbf{148} \\
sigstore/rekor          & \textbf{-0.0505} & \textbf{0.951} & \textbf{-4.92} & \textbf{$2.63{\times}10^{-17}$} & \textbf{186} \\
\midrule
keyfactor/signserver-ce & 0.0126            & 1.013          & 1.27           & 0.408                           & 22            \\
keyfactor/ejbca-ce      & -0.0001           & 1.000          & -0.01          & 0.991                           & 146           \\
\midrule
notaryproject/notation  & \textbf{-0.0322} & \textbf{0.968} & \textbf{-3.17} & \textbf{$1.20{\times}10^{-11}$} & \textbf{250}  \\
openpubkey/openpubkey   & \textbf{0.0404}  & \textbf{1.041} & \textbf{4.12}  & \textbf{$3.39{\times}10^{-7}$}  & \textbf{96}   \\
hashicorp/vault         & \textbf{-0.0250} & \textbf{0.975} & \textbf{-2.47} & \textbf{$3.00{\times}10^{-37}$} & \textbf{1{,}401} \\
\bottomrule
\end{tabular}
\end{table}



\begin{figure}[t]
    \centering
    \includegraphics[trim=20bp 265bp 0bp 24bp,clip,width=\textwidth]{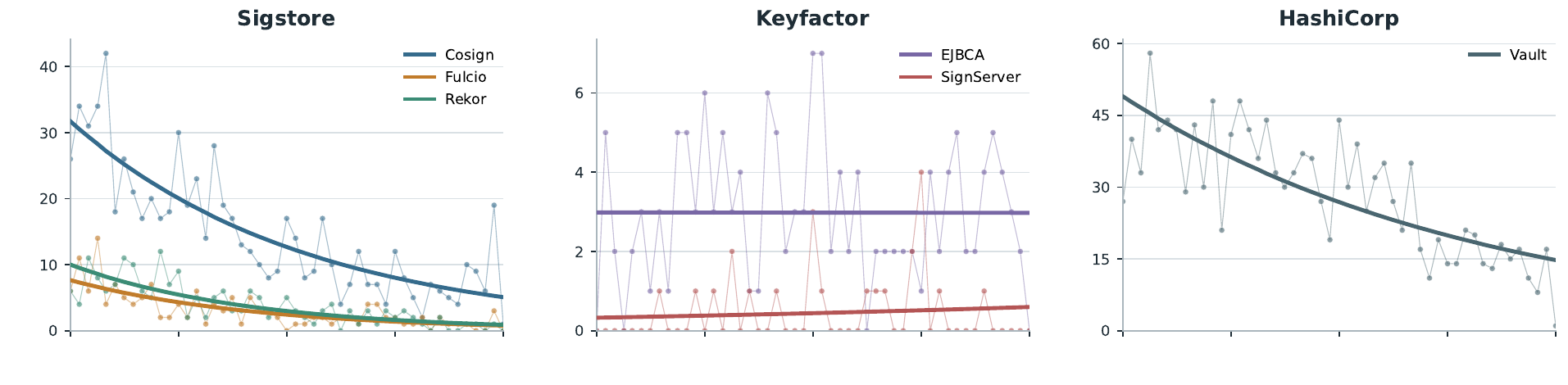}
    \par\vspace{0.25ex}
    \makebox[\textwidth][c]{\small\textit{(a) Sigstore, Keyfactor, and HashiCorp repositories.}}
    \par\vspace{0.2em}
    \includegraphics[trim=190bp 0bp 145bp 260bp,clip,width=0.7\textwidth]{fig/new/overall_poisson_observed_vs_fitted_by_ecosystem.pdf}
    \par\vspace{0.25ex}
    \makebox[0.9\textwidth][c]{\small\textit{(b) Notary and OpenPubKey repositories.}}
    \vspace{3pt}
    \caption{Observed and fitted monthly counts of usability-relevant issues, grouped by ecosystem for readability.
    Panel (a) shows Sigstore (Cosign, Fulcio, and Rekor), Keyfactor (EJBCA and SignServer), and HashiCorp (Vault); panel (b) shows Notary (Notation) and OpenPubKey.
    The lighter point-and-line series show observed monthly counts, while the thicker lines show fitted means from the repository-specific Poisson models reported in \cref{tab:poisson-overall-trends}.
    The ecosystem grouping is visual only; repositories were not aggregated.
    The five ecosystem plots use independent vertical scales to make within-ecosystem temporal patterns visible, so absolute vertical magnitudes should not be compared across ecosystems.}
    \label{fig:repository-temporal-fits}
\end{figure}

\cref{fig:repository-temporal-fits} shows the observed monthly counts alongside the repository-specific fitted Poisson trends, allowing the trajectories reported in \cref{tab:poisson-overall-trends} to be compared directly with the underlying monthly observations.

These results show that reported usability concerns do not follow a common temporal pattern across the studied tools.
The strongest declines occur in the Sigstore components, followed by Notation and Vault, while OpenPubKey moves in the opposite direction and the Keyfactor repositories show little evidence of systematic change.
This heterogeneity is consistent with the range of temporal patterns anticipated by the Tool Usability Lifecycle introduced in \cref{sec:Background:UsabilityTheory}; we return to its interpretation in \cref{sec:discussion-temporal-trajectories}.

Because five of the eight repository-level series exhibited overdispersion, where the variance in monthly counts exceeded that assumed by the Poisson model, and two exhibited significant residual autocorrelation, where residuals remained correlated across adjacent months, we treat classical Poisson $p$-values cautiously (as directional evidence) rather than relying on them as exact inferential quantities.
We therefore re-estimated the trends using HC0 heteroskedasticity-robust standard errors~\cite{white1980heteroskedasticity} and negative-binomial sensitivity~\cite{cameron_trivedi_2013} models to assess robustness to variance misspecification.
These checks preserved the direction and significance status of the trends reported in~\cref{tab:poisson-overall-trends}.
Residual autocorrelation remains a limitation of the linear time-trend specification and is considered in our interpretation of these results.
Full diagnostics and sensitivity results are reported in~\cref{sec:appendix-methodology-detail-poisson-assumption-check}.

\subsubsection{Theme- and Heuristic-Level Temporal Trends}
\label{sec:results:RQ3:tool_theme}


The repository-level (tool-component) trends summarize overall changes in reported usability issues, but they do not show which types of usability concerns account for those changes.
We therefore examine temporal trends separately for the inductive themes and Nielsen heuristics within each tool.

\begin{figure*}[t]
    \centering
    \includegraphics[trim=0 18bp 0 0,clip,width=0.99\linewidth]{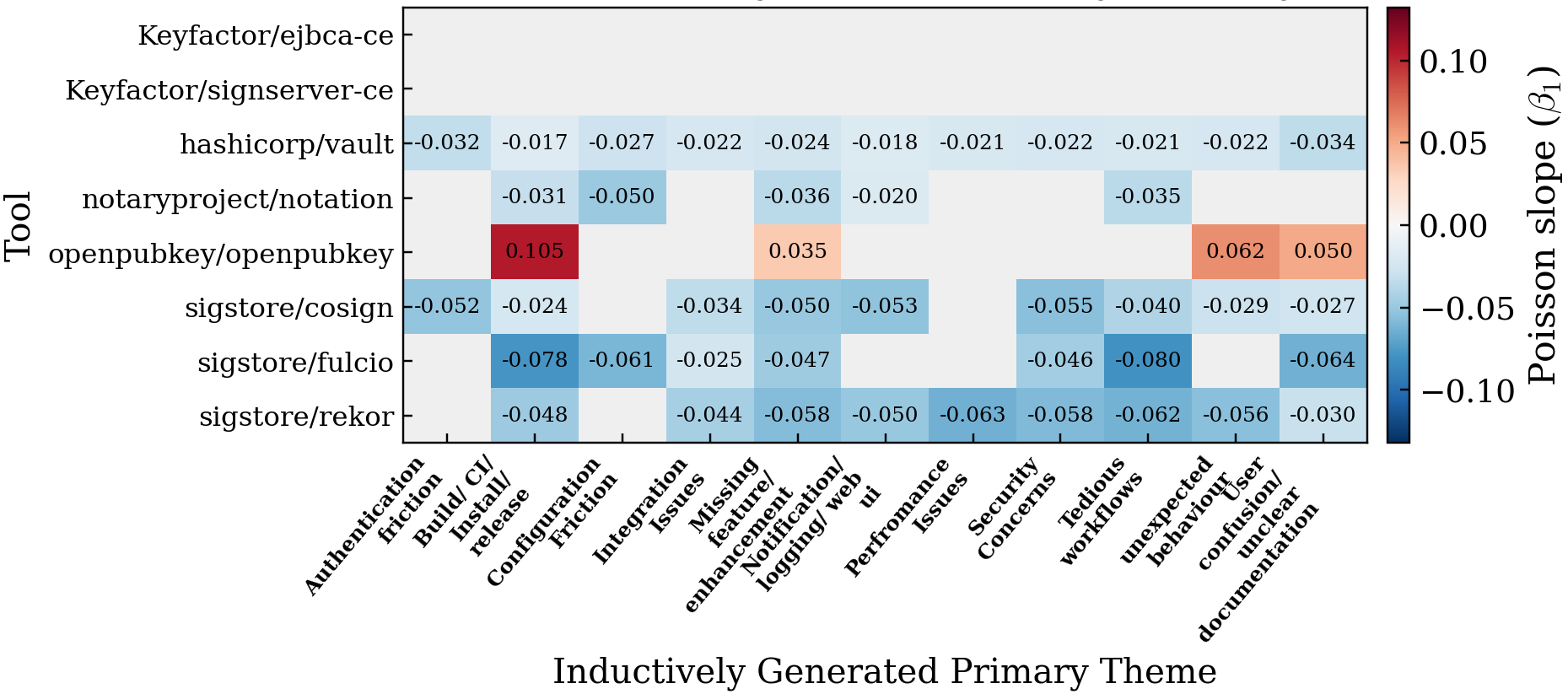}
    \par\vspace{-0.4em}{\normalsize Inductive Theme}
    \caption{    
    Heatmap of Poisson time-trend slopes ($\beta_1$) by tool and inductive theme.
    Negative values indicate decreasing expected monthly issue counts over time; positive values indicate increasing counts.
    Only statistically significant cells ($p<0.05$) are colored and annotated with the estimated $\beta_1$; non-significant cells are masked.
    }
    \label{fig:rq3-heatmap-l1-sig}
\end{figure*}

\myparagraph{Inductive Theme Trends.}

~\Cref{fig:rq3-heatmap-l1-sig} shows the estimated time trends for each tool and inductive theme.
Among the tools with overall declining trajectories, the decreases span several usability themes but are not uniform.
Vault shows significant declines across all 11 themes.
Rekor shows significant declines in 9 of 10 observed themes, Cosign in 9 of 11, and Fulcio in 7 of 11.
Notation shows significant declines in 5 of 11 themes:
\emph{Build/CI/Installation/Distribution/Release Issues},
\emph{Configuration Friction},
\emph{Missing Feature/Enhancement Request},
\emph{Notification/Logging Issues}, and
\emph{Tedious Workflows}.
At the broader theme-family level, Notation also shows a significant decline in Functional Reliability
($\beta_1=-0.0224$, $p=.028$).

OpenPubKey shows the opposite pattern, with four themes increasing significantly over the observation period.
Neither Keyfactor repository shows significant trends across the inductive themes.
These results show that the overall tool-level trajectories are composed of different changes across individual types of usability concerns.

\myparagraph{Nielsen Heuristic Trends.}

The Nielsen heuristic trends show a similar but more heterogeneous pattern (Appendix~\cref{fig:rq3-heatmap-nielsen}).
Vault shows eight significant negative heuristic trends, while Notation shows five and the Sigstore components show seven to eight each.
OpenPubKey shows four significant positive heuristic trends.
For Keyfactor, EJBCA shows one significant positive trend and SignServer shows none.

Together, the inductive and Nielsen analyses show that changes in reported usability concerns differ not only across tools but also across types of usability problems within the same tool.

\subsubsection{Pooled Theme- and Affected-Surface Temporal Trends}
\label{sec:results:RQ3:pooled}

\begin{figure*}[t]
    \centering
    \includegraphics[width=0.94\textwidth,trim=0 0 0 37bp,clip]{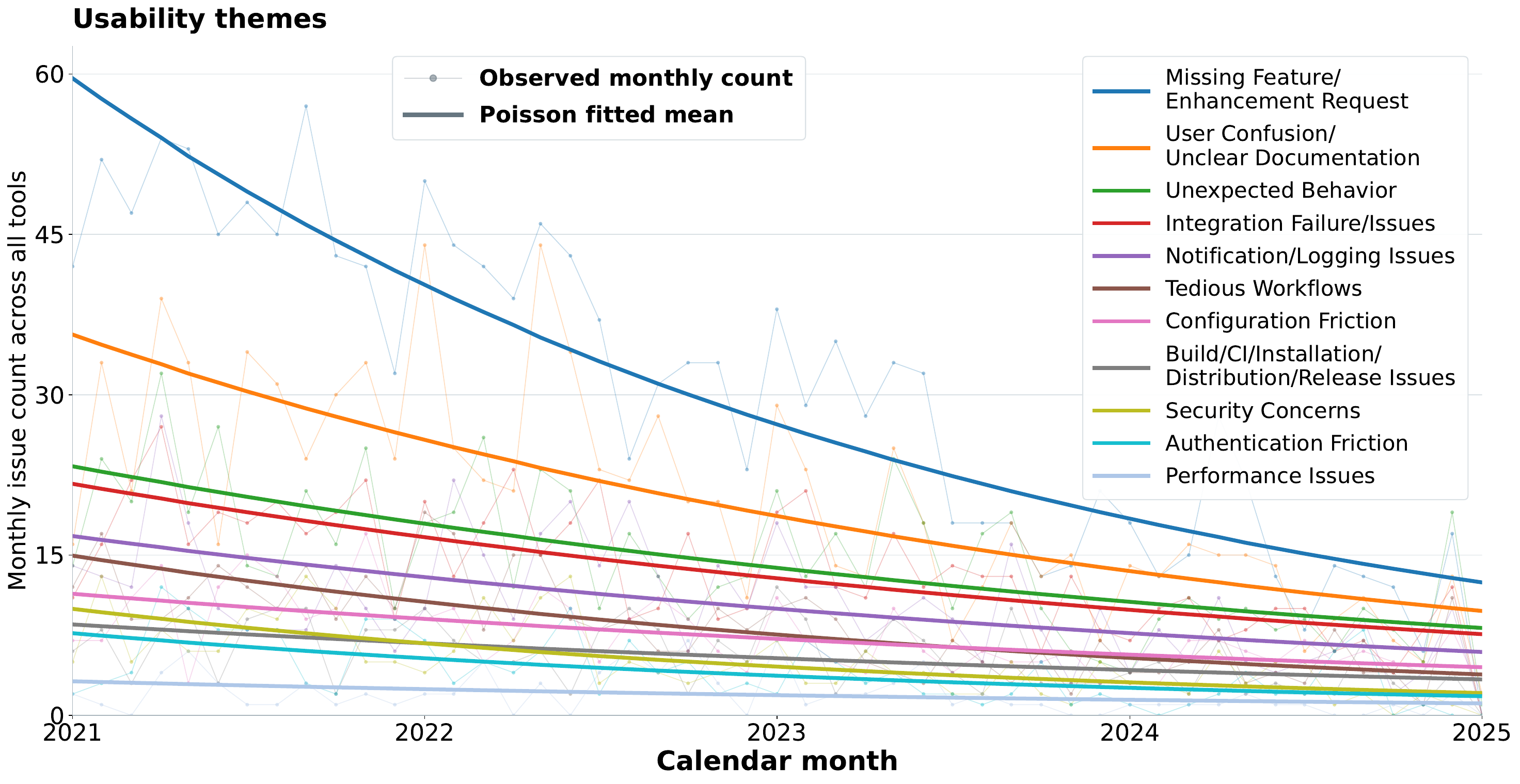}
    \par\vspace{0.2ex}
    \makebox[0.94\textwidth][c]{\small\textit{(a) Observed monthly counts and fitted means by \textbf{usability theme}.}}
    \par\vspace{0.35em}
    \includegraphics[width=0.94\textwidth,trim=0 0 0 37bp,clip]{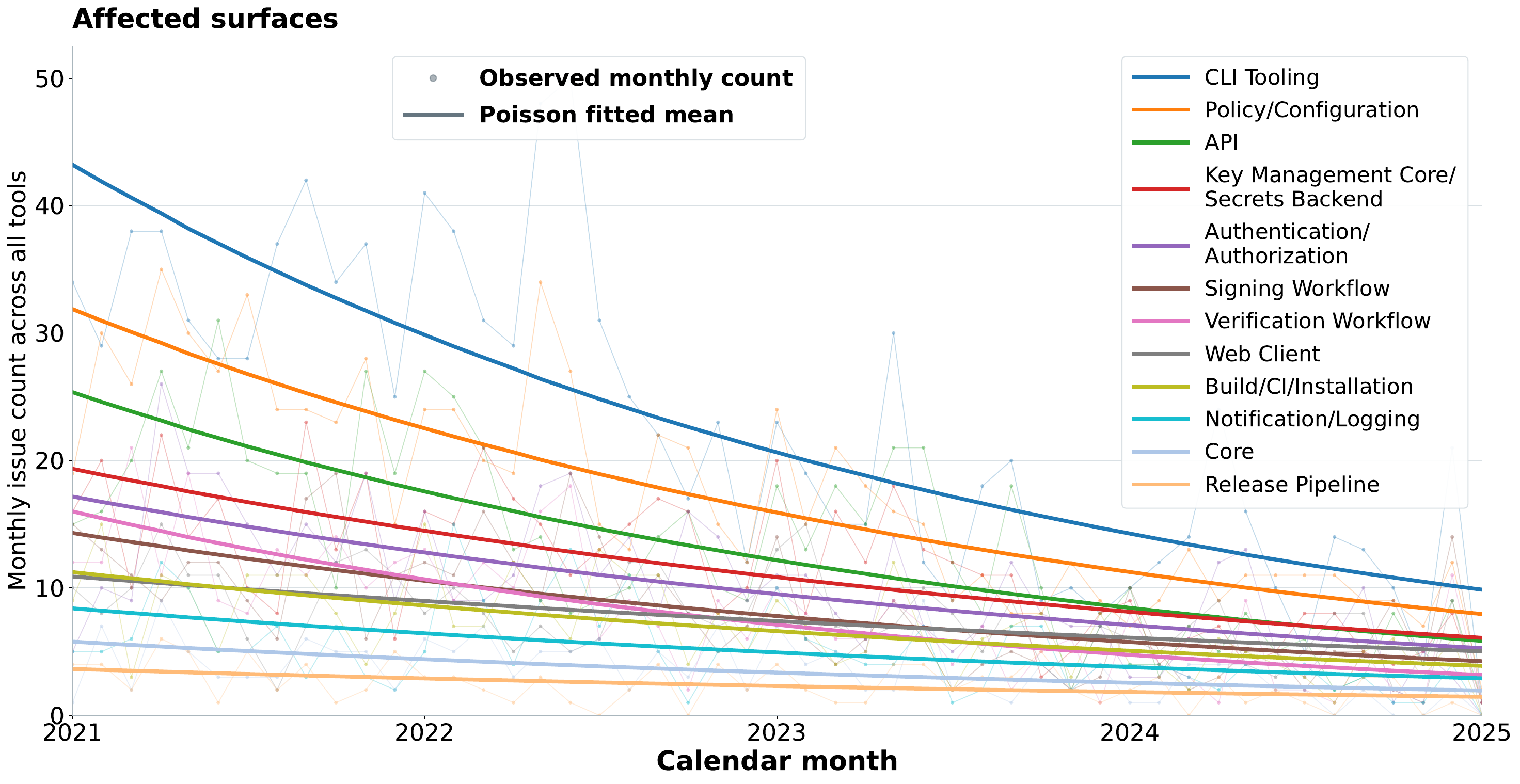}
    \par\vspace{0.2ex}
    \makebox[0.94\textwidth][c]{\small\textit{(b) Observed monthly counts and fitted means by \textbf{affected surface}.}}
    \vspace{0.03cm}
    \caption{Observed and fitted aggregate monthly counts of usability-relevant issues across the studied tools from November 2021 through November 2025. Panel (a) groups counts by usability theme, and panel (b) groups counts by affected surface. Thin point-and-line series show observed monthly counts, while thick curves show fitted Poisson means. Because the coding is multi-label, an issue may contribute to multiple series. Panels use independent vertical scales. The overlays provide a descriptive comparison of observations and fitted means; formal model diagnostics and sensitivity analyses are reported in \cref{sec:appendix-methodology-detail-poisson-assumption-check}.}
    \label{fig:rq3-aggregate}
\end{figure*}

To complement the tool-level models, we pool issue counts across tools and fit Poisson regressions by theme and by affected surface.
\Cref{fig:rq3-aggregate} shows the observed monthly counts alongside the corresponding fitted Poisson means: panel (a) summarizes trends by usability theme (\emph{what}), while panel (b) summarizes trends by affected surface  (\emph{where}).
The lighter traces show observed monthly counts, while the fitted curves summarize the estimated temporal direction of each category.
Because the Poisson mean is modeled as $\exp(\beta_0+\beta_1t)$, negative slopes indicate declines in expected monthly issue counts, while positive slopes indicate increases.

At the inductive-theme level, fitted aggregate curves decline across all modeled themes.
Estimated slopes are consistently negative, ranging from about $-0.0193$ to $-0.0326$  (Appendix~\cref{tab:poisson-aggregate-l1-theme}).
The steepest declines are observed for \emph{Missing Feature/Enhancement Request} ($\beta_1 \approx -0.0326$), \emph{Security Concerns} ($\beta_1\approx -0.0326$), \emph{Authentication Friction} ($\beta_1\approx -0.0304$), and \emph{Tedious Workflows} ($\beta_1\approx -0.0284$).
Other themes decline more gradually, including \emph{Build/CI/Installation/Distribution/Release Issues} and \emph{Configuration Friction} (both $\beta_1\approx -0.0193$).
Thus, reported issue counts decline across all modeled inductive themes in the corpus, although the rate of decline varies across problem types.

At the affected-surface level, aggregate fitted curves are also uniformly downward in the current dataset.
All modeled affected-surface slopes are negative (Appendix~\cref{tab:poisson-aggregate-component}).
The steepest declines are observed for \emph{Verification Workflow} ($\beta_1\approx-0.0338$) and CLI Tooling ($\beta_1\approx-0.0308$), followed by \emph{API} and \emph{Policy/Configuration}.
Other affected surfaces, including \emph{Authentication/Authorization}, \emph{Web Client}, and \emph{Build/CI/Installation}, also show declining reported issue counts over time.
Thus, the pooled corpus shows decreasing issue counts across all modeled affected surfaces, although the rate of decline differs across surfaces.

Across both analyses, the observed monthly counts fluctuate around the fitted trajectories, so the fitted curves should be interpreted as summaries of the overall temporal direction rather than as descriptions of month-to-month change.
All 23 aggregate slopes (11 themes and 12 affected surfaces) remained negative and statistically significant under both HC0 robust inference and negative-binomial sensitivity models; full diagnostics and robustness analyses are reported in \cref{sec:appendix-methodology-detail-poisson-assumption-check}.

These aggregate declines should be interpreted alongside the differences observed across individual tools.
A downward trend in the pooled corpus can coexist with increases for individual tools, such as OpenPubKey, because the pooled models combine repositories with different issue volumes and temporal patterns.
Thus, the pooled models describe the overall direction of reported usability concerns across the corpus, while the tool-level models show where that pattern differs across individual tools.
Because themes and affected surfaces decline at different rates, their relative prominence in the corpus also changes over time.

\section{Discussion and Future Work}
\label{sec:discussion}

\subsection{Usability Lessons for Identity-Based Signing}
\label{sec:discussion_arch_temporal}

We distill three lessons from our study.

\subsubsection{Identity-Based Signing Shifts Rather Than Eliminates Usability Burden.}

Across RQ1 and RQ2, usability remains a prominent hurdle within the studied identity-based signing ecosystems (\cref{sec:results:RQ1,sec:results_rq2}).
Our findings reinforce prior evidence that modern, identity-centric approaches have not entirely solved the usability problem~\cite{kalu2025johnnysignsnextgenerationtools}.
In particular, substantial challenges remain around configuration, integration, verification, feedback, and missing workflow capabilities.

Identity-based signing reduces much of the direct key-management burden associated with earlier cryptographic approaches~\cite{whitten1999johnny}.
However, reducing this burden does not eliminate usability challenges; rather, the burden shifts toward coordinating identity providers, policy configuration, automated infrastructure, and verification workflows.
Users perform fewer long-lived key-management tasks but must now reason about complex, security-relevant information, such as identity claims, issuers, certificates, and trust policies, to interpret verification outcomes.

This pattern is consistent with Goldratt's Theory of Constraints, which suggests that once a dominant system constraint is alleviated, other constraints can become more salient~\cite{anderson2003agile}.
In identity-based signing, reducing the burden of long-lived key management may therefore expose usability constraints in identity configuration, policy specification, cross-component coordination, and verification workflows.
This interpretation also echoes Tesler's Law of Conservation of Complexity, which suggests that reducing complexity in one part of an interaction may shift some of that complexity elsewhere~\cite{yablonski2024laws}.

\subsubsection{Usability Friction Extends Across the Signing Ecosystem}

In the studied ecosystems, reported usability problems are not evenly distributed across affected surfaces.
CLI tooling, policy/configuration, and verification workflows are among the most frequently implicated surfaces in issue reports (\cref{tab:rq2-components-top}).
Together, these findings show that usability friction appears both at direct interaction surfaces and at the boundaries where identity, policy, and workflow components must operate in tandem.

Sigstore illustrates this boundary-level friction: verification and configuration concerns often involve interpreting failures, reconciling identity assertions with policy expectations, and understanding interactions among Cosign, Fulcio, and Rekor.
Vault provides a complementary case, where usability concerns around policy and configuration arise within broader organizational workflows spanning authorization rules, backend configuration, and CI/CD integration.

These examples reinforce a broader pattern across the studied tools: usability problems extend beyond individual interfaces to the coordination points among components.
Architectural modularity enables a separation of concerns, but it also creates additional surfaces that users must understand and configure correctly.
Identity-based signing distributes responsibility across signing clients, identity providers, certificate authorities, transparency logs, policy configurations, and downstream verification workflows.
Usability therefore depends not only on whether each component works in isolation, but also on whether users can understand and manage the interactions among them.
Verification deserves particular attention here: when failures span identity, certificate, policy, or logging components, unclear feedback can make it difficult for users to identify which part of the trust chain caused verification to fail.


\subsubsection{Usability Concerns Follow Different Temporal Trajectories}
\label{sec:discussion-temporal-trajectories}

Our longitudinal results demonstrate that reported usability concerns do not follow a uniform temporal trajectory across identity-based signing ecosystems or concern types (\cref{tab:poisson-overall-trends,fig:repository-temporal-fits}).
For instance, while Sigstore, Vault, and Notation exhibit significant overall declines in reported usability issues, the Keyfactor repositories show little systematic change, and OpenPubKey exhibits an increasing trajectory.
Furthermore, our theme and affected-surface analyses reveal that even within a single tool, specific usability concerns can change at different rates (\cref{fig:rq3-heatmap-l1-sig,fig:heatmap-beta-associated_component_theme}).

These divergent patterns align with the \emph{Tool Usability Lifecycle} introduced in \cref{sec:Background:UsabilityTheory}.
As an ecosystem evolves, broader adoption and integration into new contexts can expose previously unreported problems, while continued development can address recurring friction through improved defaults, documentation, and workflow support.
The balance between these two processes provides one explanation for the heterogeneous trajectories we observe.
Consequently, usability evolution cannot be reduced to a single tool-level trajectory.
The pooled corpus shows broad declines, yet individual tools, and individual concern types within those tools, can follow different or even opposing trajectories.
Looking only at overall issue volume can therefore mask which usability constraints remain prominent over time.

We emphasize that these findings do not establish that declining issue counts inherently indicate improved usability or greater tool maturity.
Issue trajectories can also be influenced by changes in adoption, project activity, reporting practices, and the populations using a tool.
Rather, our findings show that publicly reported usability friction evolves unevenly across tools and problem types, underscoring the value of longitudinal analyses that examine both aggregate and disaggregated trajectories.



\subsection{Contrast with Prior Usability Work}

Prior empirical studies of identity-based software signing characterize usability concerns at a broader level, including adoption barriers, workflow integration challenges, documentation gaps, and coordination complexity across ecosystem components~\cite{usenix_2025_signing_interview_kalu,kalu2025johnnysignsnextgenerationtools}.
Our results complement and extend these findings in two ways.

First, our analysis provides a more granular view of these concerns.
Where prior studies identify integration complexity or documentation gaps, we characterize the specific usability themes involved, identify the affected surfaces where they manifest, and quantify their relative prominence across our large issue corpus.
This provides concrete evidence about where recurring usability friction appears within signing workflows and where engineering intervention may be most useful.

Second, we examine these concerns across five identity-based signing ecosystems rather than a single ecosystem.
The same broad usability concerns recur across tools, but their prevalence and temporal trajectories vary across tools and affected surfaces.
This cross-tool, longitudinal view therefore complements the richer point-in-time accounts provided by interview studies.

Prior usability research on cryptographic libraries and adjacent secure-communication tools commonly emphasizes safer APIs, clearer documentation, better defaults, and reducing exposure to confusing low-level primitives~\cite{green_developers_2016,Patnaik_Hallett_Rashid,whitten1999johnny}.
Much of this work focuses on usability problems at comparatively bounded tool or API interfaces.
Our findings extend the perspective provided by this prior research by showing that the usability boundary of identity-based signing frequently spans multiple interacting components and workflows.
A signing or verification problem may involve the signing client, identity provider, certificate authority, transparency log, policy configuration, or the interfaces among them.

\begin{table*}[t]
\centering
\caption{Design priorities derived from our empirical findings and connections to broader software-engineering practices.
The agentic-engineering implication is forward-looking and was not directly evaluated in this study.}
\label{tab:discussion-design-priorities}
\small
\footnotesize
\begin{tabularx}{\textwidth}{
    >{\raggedright\arraybackslash}p{0.22\textwidth}
    >{\raggedright\arraybackslash}p{0.16\textwidth}
    >{\raggedright\arraybackslash}X
    >{\raggedright\arraybackslash}p{0.22\textwidth}}
\toprule
\textbf{Empirical Observation} &
\textbf{Design Priority} &
\textbf{Implication for Signing Tools} &
\textbf{Related SE Practice} \\
\midrule

Verification failures can involve identity, certificate, policy, and logging information &
Actionable failure diagnosis &
Report the failed trust condition, relevant evidence, and a recoverable next step rather than only a generic verification failure. &
Failure-aware engineering and systematic failure learning~\cite{yin2011empirical, anandayuvaraj2022reflecting,anandayuvaraj2025learning}. \\

\addlinespace

Usability friction frequently appears at boundaries among signing components and surrounding infrastructure &
Cross-component observability &
Expose enough state to trace identity binding, credential issuance, log interaction, policy evaluation, and verification across component boundaries. &
Observability, distributed tracing, and well-defined interfaces~\cite{oppenheimer2003internet, li2022observability,xie2025tracepicker,parnas1985modular}. \\

\addlinespace

Policy/configuration is a prominent affected surface and governs downstream trust decisions &
Pre-deployment policy validation &
Provide policy validators, configuration linters, dry-run modes, and tests that expose invalid or inconsistent trust assumptions before deployment. &
Configuration testing and misconfiguration analysis~\cite{yin2011empirical, xu2016early, halin2019test,liu2026rethinking}. \\

\addlinespace
\midrule

Trust requirements are distributed across policy, identity, workflow, and infrastructure artifacts &
Machine-checkable trust obligations for agentic engineering &
Externalize consequential trust requirements and enforce settled obligations through models, constraints, validators, or gates that both humans and coding agents can inspect and use. &
Governable and model-based agentic software engineering~\cite{davis2026cheap,davis2026model}. \\

\bottomrule
\end{tabularx}
\end{table*}

These findings motivate four design priorities for identity-based signing ecosystems, summarized in \cref{tab:discussion-design-priorities}:

\begin{enumerate}[leftmargin=*, label=\textbf{\arabic*.}, itemsep=3pt, topsep=2pt]

    \item \textbf{Actionable Failure Diagnosis.}
    Verification failures should identify the failed trust condition, surface relevant evidence, and provide a recoverable next step rather than only report that verification failed.
    This priority is motivated by the prominence of N9 (\emph{Help Users Recognize, Diagnose, and Recover from Errors}) and by verification problems spanning identity, certificate, policy, and logging surfaces, and aligns with broader work on systematic learning from recurring failures~\cite{anandayuvaraj2022reflecting,anandayuvaraj2025learning}.

    \item \textbf{Cross-Component Observability.}
    When failures emerge from interactions among the signing client, identity provider, certificate authority, transparency log, and policy engine, users need enough cross-component state to reconstruct what occurred.
    This priority aligns with observability and distributed-tracing practices that expose execution evidence across component boundaries to support troubleshooting and root-cause analysis~\cite{li2022observability,xie2025tracepicker}.

    \item \textbf{Pre-Deployment Policy Validation.}
    Because Policy/Configuration is a prominent affected surface, signing ecosystems should help users identify invalid or inconsistent trust assumptions before deployment through mechanisms such as policy validators, configuration linters, dry-run modes, and tests.
    This direction is consistent with prior work on configuration testing and misconfiguration arising from constraint violations and component interactions~\cite{halin2019test,liu2026rethinking}.

    \item \textbf{Machine-Checkable Trust Obligations for Agentic Engineering.}
As coding agents increasingly modify policy, configuration, CI/CD, and deployment artifacts, consequential identity-based signing requirements should be represented in forms that both humans and agents can inspect and validate.
Externalizing these requirements as engineering models may reduce the need to repeatedly reconstruct trust assumptions from repository context and allow signing and verification changes to be checked against explicit trust obligations before deployment~\cite{davis2026cheap,davis2026model}.

\end{enumerate}

Together, these priorities shift attention from isolated interfaces toward making the broader signing ecosystem diagnosable, observable, testable, and governable.

Beyond these design priorities, the current software supply chain regime is also being reshaped by agent-mediated development, which may change what identity-based signing is expected to establish.
Identity-based signing binds an artifact to an authenticated signing identity, but increasingly agent-generated code can separate the identity that authorizes or releases an artifact from the entity that produced substantial portions of its implementation.
This distinction raises questions about whether provenance should capture only the authenticated signer or also the human--agent context through which an artifact was produced.
As agentic development changes how software is generated, reused, and integrated, the role of identity-based signing may therefore need to evolve alongside the provenance guarantees expected from it~\cite{singla2026software}.

\subsection{Methodological Implications for Studying Usability in Software Engineering}
One methodological implication of our study is that repository issue discussions can complement traditional usability methods when studying infrastructure-oriented software systems over time.
Usability research commonly relies on controlled evaluations, interviews, surveys, or direct observation of users.
These approaches provide rich evidence about user experience, but repeating them longitudinally can be difficult, particularly for infrastructure tools whose usability concerns emerge through interactions with CI/CD systems, identity providers, policy configurations, and other parts of a larger software ecosystem.
Repository issue discussions provide a complementary longitudinal record of usability problems encountered during development, deployment, and maintenance workflows.


A second methodological implication concerns combining complementary analytical perspectives.
We combine inductively generated themes, deductive Nielsen heuristic mappings, and affected-surface annotations to provide three views of a reported usability concern: \emph{what} domain-specific problem is being reported, how that problem relates to established usability principles, and \emph{where} in the software ecosystem the problem manifests.
This combination connects usability analysis with software-engineering concerns about components, interfaces, integrations, and workflows.
It may therefore be useful for studying other developer-facing systems whose usability depends on interactions among multiple tools and components rather than on a single interface.

A third methodological implication concerns the granularity at which longitudinal usability trends are analyzed.
The \emph{Tool Usability Lifecycle} (\cref{sec:Background:UsabilityTheory}) provides an interpretive framework for examining whether reported usability concerns increase, remain relatively stable, or decline as tools evolve.
Our results show, however, that a single repository-level or corpus-level trajectory can mask different temporal patterns across particular usability concerns and affected surfaces.
Future longitudinal usability studies should therefore examine trends at multiple levels, including overall tool trajectories as well as more disaggregated trends by problem type and system surface.

Finally, repository-based longitudinal evidence should be interpreted as a complement to, rather than a replacement for, direct usability measurement.
Changes in GitHub issue counts can reflect changes in adoption, reporting practices, project activity, or user populations in addition to changes in usability.
The value of this approach is therefore not that issue trajectories directly measure usability improvement, but that they provide observable longitudinal signals of where reported friction persists, declines, or emerges.
These signals can help identify targets for subsequent qualitative, observational, or user-centered investigation.


\section{Threats to Validity}
\label{sec:threats_to_validity}

\subsection{Data and Construct Validity}
\label{sec:threats:data-construct}

Data validity concerns whether the observed data adequately represent the phenomenon under study, while construct validity concerns whether our operationalizations adequately capture the intended concepts.

Our data consist of publicly reported GitHub issue discussions and therefore capture only one observable channel of usability friction.
Other feedback channels, including enterprise support, telemetry, internal monitoring, mailing lists, and internal team discussions, may capture concerns that do not appear in public repositories.
Reporting practices may also differ across projects because of governance structure, organizational backing, and user-population norms; for example, some enterprise or government deployments may triage problems through private channels.
Our dataset should therefore be interpreted as a partial view of publicly reported usability friction rather than a comprehensive measure of ecosystem usability.

A notable feature of our corpus is that 76.0\% of the filtered issue discussions were classified as usability-relevant.
This should not be interpreted as a population-level usability rate because the denominator consists of publicly reported issue discussions rather than individual users, usage events, or deployments.
However, the prominence of usability-relevant concerns is consistent with observations from other complex software systems, where substantial portions of reported problems arise at semantic, configuration, operator-interaction, and component-boundary layers rather than from low-level crashes alone~\cite{zhu2026agentframeworksfallshort,oppenheimer2003internet,yin2011empirical,xu2016early}.
Although these studies examine different domains and use different problem taxonomies, they provide useful context for why complex infrastructure software may exhibit issue profiles dominated by higher-level coordination, configuration, and interaction concerns.
Identity-based signing shows a related pattern, with substantial reported friction concentrated around configuration, integration, verification, feedback, and workflow support.

For construct validity, we operationalize usability through our usability-relevance criteria, inductively generated themes, deductive Nielsen heuristic mappings, and affected-surface annotations.
Issues were classified using LLM-assisted coding grounded in a human-developed coding scheme and evaluated against held-out human annotations.
Although this calibration and validation reduce classification error, some misclassification remains possible.
Affected-surface annotations likewise simplify interactions that may span multiple components and system boundaries.

Repository discussions also do not always reveal the underlying cause of a reported problem.
A user may frame a difficulty as an interface or workflow problem when the deeper constraint is architectural or technical, a phenomenon related to the ``\textit{XY Problem}''~\cite{WooledgeXyProblem}.
Conversely, an issue initially reported as a technical defect may reveal a usability barrier through subsequent discussion.
Repository artifacts therefore make it difficult to distinguish user-facing symptoms from deeper architectural constraints or design trade-offs in every case.
Consequently, our findings should be interpreted as patterns in publicly reported usability concerns rather than definitive diagnoses of underlying technical root causes.

\subsection{Internal Validity} 
\label{sec:internal_validity}

Internal validity concerns alternative explanations for the observed patterns.
Our analyses characterize associations and temporal patterns in publicly reported usability concerns rather than causal changes in usability itself.
Temporal trends in issue counts may also reflect changes in adoption, project activity, release cadence, reporting practices, or the populations using a tool.
Similarly, differences across tools may reflect ecosystem size, maturity, governance, or organizational backing in addition to differences in reported usability friction.
We therefore avoid interpreting declining issue counts as direct evidence of usability improvement or tool maturity, and instead interpret the longitudinal results as changes in the volume of publicly reported usability concerns during the observation period.

\subsection{External Validity}
\label{sec:external_validity}

External validity concerns how well our findings generalize beyond the studied ecosystems.
Our dataset comprises publicly reported GitHub issue discussions from eight repositories representing five prominent identity-based signing ecosystems with public development repositories.
The findings may not generalize to proprietary signing systems, ecosystems with different issue-reporting practices, or deployment contexts in which usability concerns are handled through private support, telemetry, internal monitoring, or organizational communication channels.
For example, Sigstore is used in commercial and government settings, including by the U.S. Department of Defense~\cite{Seagren2022SopsSigstore}, and those experiences may not be fully represented in public repositories.
Our results should therefore be interpreted as evidence about publicly observable usability concerns in the studied ecosystems rather than as estimates of usability across all identity-based signing deployments.

\subsection{Statistical Conclusion Validity} 
\label{sec:statistical_conclusion_validity}

This section addresses the reliability of quantitative inferences.
Our longitudinal analyses rely on Poisson regression, and clustering. Overdispersion or temporal dependence could affect estimates.
We interpret statistical results as directional evidence rather than precise effect magnitudes.
Pearson dispersion statistics across the eight repositories ranged from 0.892
(Rekor) to 7.156 (OpenPubKey), with five of eight repositories exceeding the overdispersion threshold of 1.5.
Residual autocorrelation was significant for Notation and OpenPubKey, indicating
temporal clustering not fully captured by a single linear slope.

For the eight overall repository-level trends, we addressed these concerns in two ways:
(1) re-estimating the models with HC0 heteroskedasticity-robust standard errors; and
(2) fitting negative-binomial sensitivity models.
Both robustness checks preserved the direction and significance status of all eight overall repository-level trends.
The principal caveat is OpenPubKey, where the estimated magnitude
of the positive trend is sensitive to model choice
($\beta_\text{Poisson} = 0.040$ vs.\ $\beta_\text{NB} = 0.104$), although
the positive direction and statistical significance remain consistent across both specifications.

Taken together, these robustness analyses indicate that the central repository-level and aggregate temporal results are stable across alternative variance specifications, with the direction and significance status preserved under HC0 robust inference and negative-binomial sensitivity models.
For the more highly disaggregated tool-by-theme, tool-by-theme-family, tool-by-Nielsen, and tool-by-affected-surface analyses, we retain the classical Poisson estimates as exploratory analyses and emphasize direction, magnitude, uncertainty, and consistency across related estimates rather than isolated $p$-values.
Full diagnostics and robustness analyses are reported in
\cref{sec:appendix-methodology-detail-poisson-assumption-check,tab:poisson-diagnostics-summary,tab:poisson-diagnostics-overall-by-tool,tab:poisson-overall-trends-robust-se,tab:nb-sensitivity-overall-by-tool}.


\section{Conclusion}
\label{sec:conclusion}

\textit{Identity-based signing does not eliminate usability burden; it redistributes and reshapes it across system boundaries and over time.}
Identity-based signing was introduced in part to alleviate the long-standing usability challenges of key-managed cryptographic workflows.
Through a cross-tool, longitudinal analysis of approximately 3{,}900 GitHub issue discussions across eight repositories representing five identity-based signing ecosystems, we examine what usability concerns are publicly reported, where they manifest, and how their reported frequency changes over time.
Identity-based signing represents a meaningful architectural advance over traditional key management.
Yet, our results indicate that usability remains a substantial and evolving concern in secure software supply chains.
Reported problems are especially prominent around CLI tooling, policy/configuration, and verification workflows, including points where orchestrating clients, identity binding, policy evaluation, logging infrastructure, and deployment workflows intersect.
Reported usability concerns also follow different temporal trajectories across repositories, themes, and affected surfaces rather than exhibiting a uniform pattern of change.
For \textbf{tool designers and maintainers}, these findings motivate treating verification clarity, cross-component observability, policy and configuration validation, and deployment integration as first-class design targets rather than peripheral integration details.
For \textbf{researchers}, our findings underscore the value of ecosystem-level and longitudinal analyses of security-tool usability, particularly analyses that distinguish aggregate trends from changes in specific problem types and system surfaces.
Future signing systems should therefore pair strong trust models with deliberate attention to the coordination surfaces through which users configure, understand, and operationalize them.

\ifARXIV
\section{Ethics Statement}
\fi
\label{sec:appendix_ethics}

We describe our stakeholder-based ethics analysis, following the guidance of Davis \etal~~\cite{davis2025guide}.

\myparagraph{Stakeholders:}
The direct stakeholders in this study are maintainers and contributors of the signing tools whose issue trackers and repositories we analyzed.
Indirect stakeholders include downstream developers and organizations that adopt these tools, and the cybersecurity community that may interpret or act upon our findings.
The research team is also a stakeholder: we bear responsibility for accurate representation and careful interpretation of public artifacts, and the consequences of dissemination.

\myparagraph{Potential Harms:}
The primary potential harm of our work is \textit{reputational}.
We analyzed only publicly available artifacts from public software-development repositories and issue trackers.
These projects maintain public repositories to promote transparency, collaboration, and external scrutiny.
All engineers whose artifacts appear in our dataset contributed those materials in public forums.
No private communications (or data), or personally sensitive information were accessed.
Nevertheless, our findings could be interpreted as reflecting negatively on specific organizations or engineers.
To mitigate this risk, we report aggregated results at the ecosystem and component level. 

\myparagraph{Potential Benefits:}
This work aims to improve understanding of usability challenges in identity-based signing systems.
By identifying recurring friction points across multiple ecosystems, we seek to inform tool designers, maintainers, and adopters about areas where usability improvements may strengthen security outcomes.
Adopters and the broader public may benefit if improved usability increases the adoption and correct use of signing technologies.

\myparagraph{Decision to Proceed and Publish:}
We judge that the societal benefits of understanding usability barriers in security tooling outweigh the limited reputational risks of aggregate reporting.
We therefore proceeded with dissemination. 

\section{Acknowledgments}
We acknowledge support from Google, Cisco, and NSF \#2229703.

\section{Data Availability}
\label{sec:data_availability}
Our study artifacts are available at \url{https://github.com/nextgenusability/longitudinal_usability_id_signing/tree/revised}.
The artifact repository includes the final classifier prompt, coding codebook, LLM output-processing scripts, table- and figure-generation scripts, robustness-check scripts, and derived result files used to produce the manuscript tables and figures.
\bibliographystyle{template/ACM-Reference-Format}
\bibliography{references/reference}

\appendix

\section*{Outline of Appendices}

\noindent
The appendix contains the following material:

\begin{itemize}[leftmargin=12pt, rightmargin=5pt]

\item \cref{sec:bg_trad_signing}: Extended Background on traditional key-based signing (\cref{sec:appendix_bg_traditional}).
\item \cref{sec:appendix-methodology-detail}: Methodological details, including corpus context metadata (\cref{sec:appendix_corpus_context_metadata}), the codebook (\cref{sec:appendix_codebook}), and Poisson assumption checks and robustness analyses (\cref{sec:appendix-methodology-detail-poisson-assumption-check}).
\item \cref{sec:appendix_other_results}: Additional results for RQ2 (\cref{sec:appendix_results_rq2}) and RQ3 (\cref{sec:appendix_results_rq3}).
\end{itemize}

\section{Extended Background}
\label{sec:bg_trad_signing}

Here we provide additional background on traditional key-managed software signing for comparison with identity-based signing.

\subsection{Traditional Key-Based Signing}
\label{sec:appendix_bg_traditional}
\begin{figure}[h!]
    \centering
    \includegraphics[width=0.6\linewidth]{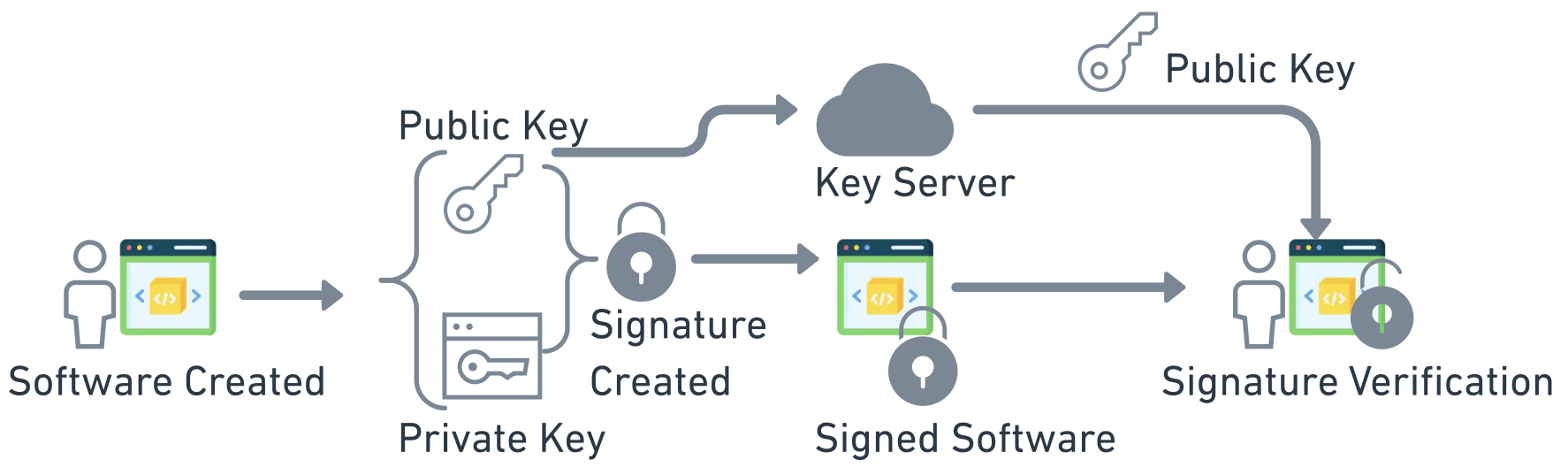}
    \caption{
    Traditional key-managed signing centers on a long-lived key pair that the signer creates and protects and that verifiers must obtain to validate signatures.
    The workflow highlights two recurring usability burdens:
public-key distribution (often via a key server)
and
establishing confidence that the retrieved public key corresponds to the intended issuer identity.
Compare this flow to that of identity-based signing in~\cref{fig:bg_id_workflow}.
    }
    \label{fig:bg_trad_workflow}
\end{figure}

Key-managed (legacy) software signing tools (illustrated in \cref{fig:bg_trad_workflow})—such as OpenPGP-style workflows—rely on long-lived public-private key pairs that are created and maintained by the signer~~\cite{openpgp_new}.
A typical workflow has three phases.
First, the signer generates a key pair and protects the private key.
Second, the signer distributes the public key (or certificate) so that verifiers can discover it.
Third, verifiers retrieve the artifact, retrieve the signer's public information, and validate the signature before trusting the artifact.
This key-managed design makes usability hinge on key lifecycle tasks that are external to the signing act itself.
Examples include key generation choices, backups, rotation, revocation, publication to \emph{key servers}, and establishing trust relationships.
Usable security research shows that these steps are difficult to execute correctly, and difficulty can reduce adoption or lead to incorrect verification decisions~~\cite{whitten1999johnny, green_developers_2016}.

\section{Methodological Details}
\label{sec:appendix-methodology-detail}

\subsection{Codebook}
\label{sec:appendix_codebook}
\cref{tab:phase3-theme-codebook} presents our codebook.

\begingroup
\scriptsize
\small
\footnotesize

\setlength{\LTleft}{0pt}
\setlength{\LTright}{0pt}
\setlength{\tabcolsep}{3pt}
\begin{longtable}
{p{0.14\textwidth} p{0.22\textwidth} p{0.48\textwidth} p{0.14\textwidth}}
\caption{Final coding scheme with theme, theme-family, heuristic, and affected-surface definitions.}\label{tab:phase3-theme-codebook}\\
\hline
\textbf{Coding Dimension} & \textbf{Label} & \textbf{Definition} & \textbf{Example Issues} \\
\hline
\endfirsthead
\hline
\textbf{Coding Dimension} & \textbf{Label} & \textbf{Definition} & \textbf{Example Issues} \\
\hline
\endhead
\hline
\endfoot
\hline
\endlastfoot
Theme & Missing feature / enhancement request & User requests functionality that does not currently exist, or enhancement of existing capability. & \href{https://github.com/Keyfactor/ejbca-ce/issues/473}{EJBCA-473} \\
Theme & Unexpected behavior & Tool behaves incorrectly despite user following expected instructions or documentation. & \href{https://github.com/sigstore/fulcio/issues/1858}{Fulcio-1858} \\
Theme & Authentication friction & User cannot authenticate/authorize smoothly due to token, flow, timeout, or identity setup problems. & \href{https://github.com/hashicorp/vault/issues/31568}{Vault-31568} \\
Theme & Configuration friction & User struggles to configure the tool due to complexity, rigid settings, or unclear setup requirements. & \href{https://github.com/hashicorp/vault/issues/31627}{Vault-31627} \\
Theme & Integration failure/issues & Interoperation with external systems fails or requires additional support/enhancement. & \href{https://github.com/notaryproject/notation/issues/1254}{Notation-1254} \\
Theme & User confusion / unclear documentation & User is confused, asks usage questions, or documentation is unclear/outdated/insufficient. & \href{https://github.com/notaryproject/notation/issues/1332}{Notation-1332} \\
Theme & Build/CI/installation/ distribution release issues & Problems running/installing/building/distributing the tool in local or CI/CD environments. & \href{https://github.com/Keyfactor/ejbca-ce/issues/824}{EJBCA-824} \\
Theme & Performance issue & Tool functions but is too slow or resource intensive (or needs performance improvements). & \href{https://github.com/Keyfactor/ejbca-ce/issues/211}{EJBCA-211}  \\
Theme & Security concerns & Reported vulnerability, unsafe default, leakage risk, or request to harden security posture. & \href{https://github.com/Keyfactor/ejbca-ce/issues/854}{EJBCA-854}  \\
Theme & Notification/Logging Issues & Logs, errors, status output, or web/client feedback are unclear, noisy, or not actionable. & \href{https://github.com/Keyfactor/ejbca-ce/issues/430}{EJBCA-430} \\
Theme & Tedious Workflows & Workflow is overly manual or requires too many steps to achieve a routine task. & \href{https://github.com/Keyfactor/ejbca-ce/issues/275}{EJBCA-275}  \\
\midrule
Theme family & Operational Friction & Barriers in setup, environment, integration, or workflow execution where users know what to do but the system resists. & \href{https://github.com/sigstore/rekor/issues/2575}{Rekor-2575} \\
Theme family & Cognitive Friction & Barriers in understanding or mental model clarity, where users do not know how to proceed. & \href{https://github.com/sigstore/cosign/issues/3671}{Cosign-3671}\\
Theme family & Functional Reliability & Friction when expected function is undermined by failures, slowness, or security risk. & \href{https://github.com/Keyfactor/ejbca-ce/issues/952}{EJBCA-952}  \\
Theme family & Functional Gap & Tool currently lacks capability needed for the user’s intended task. & \href{https://github.com/Keyfactor/ejbca-ce/issues/943}{EJBCA-943}  \\
\midrule
Nielsen heuristic & 1. Visibility of system status & System should keep users informed with timely, clear feedback about ongoing operations. & \href{https://github.com/sigstore/cosign/issues/4438}{Cosign-4438} \\
Nielsen heuristic & 2. Match between system and the real world & Use familiar language and concepts; avoid internal jargon that conflicts with user expectations. & \href{https://github.com/notaryproject/notation/issues/1170}{Notation-1170} \\
Nielsen heuristic & 3. User control and freedom & Provide clear exits/undo paths so users can recover from unwanted states. & \href{https://github.com/sigstore/rekor/issues/2575}{Rekor-2575} \\
Nielsen heuristic & 4. Consistency and standards & Use consistent terminology/behavior and follow ecosystem conventions. & \href{https://github.com/hashicorp/vault/issues/13192}{Vault-13192} \\
Nielsen heuristic & 5. Error prevention & Prevent avoidable mistakes with checks, guardrails, and confirmations before commitment. & \href{https://github.com/hashicorp/vault/issues/31586}{Vault-31586} \\
Nielsen heuristic & 6. Recognition rather than recall & Reduce memory burden by making required actions/options visible in the interface/help. & \href{https://github.com/hashicorp/vault/issues/31612}{Vault-31612} \\
Nielsen heuristic & 7. Flexibility and efficiency of use & Support efficient workflows for experts while remaining usable for less-experienced users. & \href{https://github.com/sigstore/cosign/issues/4507}{Cosign-4507} \\
Nielsen heuristic & 8. Aesthetic and minimalist design & Avoid irrelevant/noisy information that hides key outputs or decisions. & \href{https://github.com/sigstore/cosign/issues/3911}{Cosign-3911} \\
Nielsen heuristic & 9. Help users recognize, diagnose, and recover from errors & Error messages should be clear, specific, and provide actionable recovery guidance. & \href{https://github.com/hashicorp/vault/issues/31582}{Vault-31582} \\
Nielsen heuristic & 10. Help and documentation & Documentation and help text should be discoverable, accurate, and aligned with current behavior. & \href{https://github.com/sigstore/cosign/issues/3671}{Cosign-3671} \\
\midrule
Affected surface & Authentication / Authorization tools & Identity verification and permission logic, including OIDC flows, token handling, MFA, and RBAC behavior. & \href{https://github.com/sigstore/cosign/issues/4438}{Cosign-4438}
\\
Affected surface & CLI tooling & Command-line UX including command/flag behavior, prompts, and output formatting. & \href{https://github.com/notaryproject/notation/issues/1259}{Notation-1259} \\
Affected surface & Signing workflow & End-to-end client-side path used to produce cryptographic signatures for artifacts. & \href{https://github.com/notaryproject/notation/issues/1254}{Notation-1254} \\
Affected surface & Verification workflow & End-to-end client-side path used to validate signatures and trust decisions. & \href{https://github.com/sigstore/cosign/issues/3671}{Cosign-3671} \\
Affected surface & Policy/configuration & Configuration and policy expression/interpretation (e.g., YAML/JSON, OPA/Rego, policy enforcement). & \href{https://github.com/sigstore/cosign/issues/2570}{Cosign-2570} \\
Affected surface & Build/CI/Installation & Friction in installation/build and automated execution in CI/CD or scripted environments. & \href{https://github.com/notaryproject/notation/issues/1288}{Notation-1288} \\
Affected surface & Release pipeline & Maintainer-side release engineering: building, signing, and publishing official binaries/artifacts. & \href{https://github.com/Keyfactor/ejbca-ce/issues/824}{Ejbca-824} \\
Affected surface & Notification/Logging & Clarity/actionability of logs, warnings, status output, and debug traces. & \href{https://github.com/notaryproject/notation/issues/1259}{Notation-1259} \\
Affected surface & Core & Issue affects the overall tool/core behavior when a specific component boundary is unclear. & \href{https://github.com/notaryproject/notation/issues/1239}{Notation-1239} \\
Affected surface & API & Issues in API endpoints, request/response behavior, and API-facing interactions. & \href{https://github.com/hashicorp/vault/issues/31582}{Vault-31582} \\
Affected surface & Web Client & Issues in web UI/client/admin dashboard interactions and feedback. & \href{https://github.com/sigstore/cosign/issues/1108}{Cosign-1108} \\
Affected surface & Key Management Core / Secrets Backend & Key/secret lifecycle and secure storage infrastructure, including KMS/HSM/keychain integrations. & \href{https://github.com/hashicorp/vault/issues/31601}{Vault-31601} \\
\end{longtable}

\endgroup

\subsection{Poisson Assumption Checks and Robustness Analyses}
\label{sec:appendix-methodology-detail-poisson-assumption-check}

\paragraph{Poisson Assumption Checks.}
We assessed the adequacy of Poisson trend models using two dispersion diagnostics and a residual dependence check: Pearson dispersion ($\chi^2_P/\mathrm{df}$), deviance dispersion, and a lag-1 Ljung--Box test on Pearson residuals (\cref{tab:poisson-diagnostics-summary,tab:poisson-diagnostics-overall-by-tool}).
Pearson dispersion values near 1 indicate approximate Poisson variance, while values substantially greater than 1 indicate overdispersion (variance exceeding the mean).
Using a conservative flag of $\chi^2_P/\mathrm{df} > 1.5$, overdispersion was common across model sets: 5/8 overall-by-tool models, 6/11 aggregate models by inductive theme, and 7/12 aggregate models by affected surface; diagnostics for the more disaggregated models are reported in \cref{tab:poisson-diagnostics-summary}.
In the overall-by-tool fits, overdispersion was especially large for openpubkey/openpubkey ($\chi^2_P/\mathrm{df} = 7.16$) and notaryproject/notation ($4.51$), while sigstore/rekor was close to Poisson variance ($0.89$) (\cref{tab:poisson-diagnostics-overall-by-tool}).
Lag-1 residual autocorrelation was significant for notaryproject/notation and openpubkey/openpubkey, indicating temporal dependence not fully captured by a single linear time slope; no significant lag-1 residual autocorrelation was detected in the aggregate theme or affected-surface models.
These diagnostics motivate robustness checks for the central repository-level and aggregate temporal results.

\paragraph{Robust Standard Errors.}
As a first robustness check, we recomputed inference for the eight overall-by-tool Poisson models using HC0 heteroskedasticity-robust standard errors (\cref{tab:poisson-overall-trends-robust-se}).
All six statistically significant repository-level trends remained significant under HC0 inference, while both non-significant Keyfactor trends remained non-significant.
We applied the same procedure to the 11 aggregate theme and 12 aggregate affected-surface models because 6 and 7 of these series, respectively, exhibited overdispersion.
All 23 aggregate slopes remained negative and statistically significant under HC0 robust inference.
We therefore retain the classical Poisson coefficients and fitted means as the reported trend estimates while using HC0 standard errors as the primary basis for inference for these aggregate models.

\paragraph{Negative-Binomial Sensitivity.}
As a second robustness check, we fit negative-binomial trend models, which relax the Poisson variance restriction by estimating an overdispersion parameter $\hat{\alpha}$.
For the overall-by-tool models (\cref{tab:nb-sensitivity-overall-by-tool}), the direction and significance status of all eight trends were preserved.
The estimated magnitude of the OpenPubKey increase was sensitive to model choice ($\beta_\text{Poisson} = 0.040$ vs.\ $\beta_\text{NB} = 0.104$), although the direction and statistical significance were unchanged.
Negative-binomial sensitivity models likewise preserved the negative and statistically significant trends for all 11 aggregate themes and all 12 aggregate affected surfaces.
For Release Pipeline, the negative-binomial dispersion estimate approached zero, effectively reducing the model toward Poisson; although the default optimizer converged, the near-boundary solution produced an undefined standard error and $p$-value, while a documented optimizer fallback produced finite inference with the same negative and statistically significant trend.
Full negative-binomial fit diagnostics and optimizer logs are provided in the accompanying artifact.

For the more highly disaggregated tool-by-theme, tool-by-theme-family, tool-by-Nielsen, and tool-by-affected-surface analyses, we retain the classical Poisson estimates as exploratory analyses and emphasize trend direction, magnitude, and consistency rather than isolated significance tests.
HC0 and negative-binomial checks address variance misspecification but do not correct serial dependence; any residual autocorrelation identified in the diagnostics therefore remains a limitation of the linear time-trend specification.


\begin{table*}[t]
\centering
\caption{Poisson model diagnostic checks by model set (dispersion and residual autocorrelation).}
\label{tab:poisson-diagnostics-summary}
\scriptsize
\footnotesize
\begin{tabular}{llrrrrrrr}
\toprule
\textbf{Model Set} & \textbf{$N$} & \textbf{Median Disp.} & \textbf{P90 Disp.} & \textbf{$N_{>1.5}$} & \textbf{$N_{<0.8}$} & \textbf{$N_\text{autocorr}$} & \textbf{\%$_{>1.5}$} \\
\midrule
Aggregate by affected surface & 12 & 1.54 & 2.11 &  7 &  0 &  0 & 58.33 \\
Aggregate by theme            & 11 & 1.66 & 2.06 &  6 &  0 &  0 & 54.55 \\
Overall usability by tool     &  8 & 1.87 & 5.30 &  5 &  0 &  2 & 62.50 \\
Tool by affected surface      & 90 & 1.13 & 2.12 & 21 &  5 &  8 & 23.33 \\
Tool by theme                 & 86 & 1.17 & 1.86 & 22 &  8 &  9 & 25.58 \\
Tool by theme family          & 32 & 1.50 & 2.99 & 16 &  2 &  6 & 50.00 \\
Tool by Nielsen heuristic     & 77 & 1.27 & 2.24 & 22 & 11 & 10 & 28.57 \\
\bottomrule
\end{tabular}
\end{table*}


\begin{table*}[t]
\centering
\caption{Poisson diagnostic checks for overall monthly usability-count trends by tool.}
\label{tab:poisson-diagnostics-overall-by-tool}
\scriptsize
\begin{tabular}{lllllll}
\toprule
\textbf{Tool} & \textbf{Disp.\ (Pearson)} & \textbf{Disp.\ (Deviance)} & \textbf{Overdispersed $>$1.5} & \textbf{Underdispersed $<$0.8} & \textbf{Ljung-Box $p$} & \textbf{Autocorr.} \\
\midrule
Keyfactor/ejbca-ce      & 1.106 & 1.288 & no  & no & 0.453 & no  \\
Keyfactor/signserver-ce & 1.544 & 1.229 & yes & no & 0.684 & no  \\
hashicorp/vault         & 2.468 & 2.681 & yes & no & 0.780 & no  \\
notaryproject/notation  & 4.505 & 4.358 & yes & no & 0.000 & yes \\
openpubkey/openpubkey   & 7.156 & 5.324 & yes & no & 0.000 & yes \\
sigstore/cosign         & 2.202 & 1.936 & yes & no & 0.658 & no  \\
sigstore/fulcio         & 1.027 & 1.049 & no  & no & 0.165 & no  \\
sigstore/rekor          & 0.892 & 1.039 & no  & no & 0.654 & no  \\
\bottomrule
\end{tabular}
\end{table*}

\begin{table*}[t]
\centering
\caption{Sensitivity check for overall usability trends: Poisson standard errors vs.\ HC0 robust standard errors by tool. Rate ratio (RR) is per month.}
\label{tab:poisson-overall-trends-robust-se}
\scriptsize
\footnotesize
\setlength{\tabcolsep}{4pt}
\renewcommand{\arraystretch}{1.1}
\begin{tabular}{lrrrrrr}
\toprule
\textbf{Tool} & \textbf{$\beta_1$} & \textbf{SE} & \textbf{$p$} & \textbf{Robust SE} & \textbf{Robust $p$} & \textbf{RR/month} \\
\midrule
sigstore/rekor          & -0.0505 & 0.0060 & $2.63{\times}10^{-17}$ & 0.0056 & $1.69{\times}10^{-19}$ & 0.9508 \\
sigstore/fulcio         & -0.0479 & 0.0066 & $4.14{\times}10^{-13}$ & 0.0072 & $3.46{\times}10^{-11}$ & 0.9533 \\
sigstore/cosign         & -0.0382 & 0.0029 & $2.43{\times}10^{-40}$ & 0.0048 & $1.36{\times}10^{-15}$ & 0.9625 \\
notaryproject/notation  & -0.0322 & 0.0047 & $1.20{\times}10^{-11}$ & 0.0088 & $2.65{\times}10^{-4}$  & 0.9683 \\
hashicorp/vault         & -0.0250 & 0.0020 & $3.00{\times}10^{-37}$ & 0.0032 & $1.07{\times}10^{-14}$ & 0.9753 \\
Keyfactor/ejbca-ce      & -0.0001 & 0.0059 & 0.991                  & 0.0059 & 0.991                  & 0.9999 \\
Keyfactor/signserver-ce &  0.0126 & 0.0152 & 0.408                  & 0.0141 & 0.373                  & 1.0127 \\
openpubkey/openpubkey   &  0.0404 & 0.0079 & $3.39{\times}10^{-7}$  & 0.0150 & $7.20{\times}10^{-3}$  & 1.0412 \\
\bottomrule
\end{tabular}
\end{table*}

\begin{table*}[t]
\centering
\caption{Negative-binomial sensitivity check for overall monthly usability-count trends by tool.}
\label{tab:nb-sensitivity-overall-by-tool}
\scriptsize
\footnotesize
\setlength{\tabcolsep}{4pt}
\renewcommand{\arraystretch}{1.1}
\begin{tabular}{lrrrrrrrr}
\toprule
\textbf{Tool} & \textbf{$\beta_1$} & \textbf{$p$} & \textbf{RR/mo} & \textbf{NB $\beta$} & \textbf{NB SE} & \textbf{NB $p$} & \textbf{NB RR/mo} & \textbf{$\hat{\alpha}$} \\
\midrule
sigstore/rekor          & -0.0505 & $2.63{\times}10^{-17}$ & 0.9508 & -0.0506 & 0.0062 & $2.04{\times}10^{-16}$ & 0.9506 & 0.0049 \\
sigstore/fulcio         & -0.0479 & $4.14{\times}10^{-13}$ & 0.9533 & -0.0477 & 0.0068 & $2.03{\times}10^{-12}$ & 0.9535 & 0.0145 \\
sigstore/cosign         & -0.0382 & $2.43{\times}10^{-40}$ & 0.9625 & -0.0373 & 0.0038 & $4.17{\times}10^{-23}$ & 0.9634 & 0.0532 \\
notaryproject/notation  & -0.0322 & $1.20{\times}10^{-11}$ & 0.9683 & -0.0446 & 0.0112 & $6.77{\times}10^{-5}$  & 0.9564 & 0.6356 \\
hashicorp/vault         & -0.0250 & $3.00{\times}10^{-37}$ & 0.9753 & -0.0272 & 0.0033 & $6.81{\times}10^{-17}$ & 0.9731 & 0.0542 \\
Keyfactor/ejbca-ce      & -0.0001 & 0.991                  & 0.9999 & -0.0001 & 0.0061 & 0.991                  & 0.9999 & 0.0233 \\
Keyfactor/signserver-ce &  0.0126 & 0.408                  & 1.0127 &  0.0151 & 0.0206 & 0.464                  & 1.0152 & 1.1425 \\
openpubkey/openpubkey   &  0.0404 & $3.39{\times}10^{-7}$  & 1.0412 &  0.1041 & 0.0412 & $1.14{\times}10^{-2}$  & 1.1097 & 4.5421 \\
\bottomrule
\end{tabular}
\end{table*}

\section{Additional Results}
\label{sec:appendix_other_results}

\subsection{Results for RQ2}
\label{sec:appendix_results_rq2}

RQ2 asks which affected surfaces are most frequently involved in usability problems.
\cref{tab:rq2-top3-components-by-tool} shows the top-three trouble areas for each studied tool.
This table is discussed in~\cref{sec:results_rq2}.

\begin{table*}[t]
\centering
\scriptsize
\caption{
Top three affected surfaces per tool component.
Values in parentheses are within-component percentages.
\textit{Core} is excluded from the ranking as it is a cross-cutting label applied
when no specific affected surface can be identified, rather than a discrete
surface category. Where \textit{Core} would have appeared in the top three,
the next specific surface is shown instead and its displaced rank is noted.$^{\dagger}$
}
\label{tab:rq2-top3-components-by-tool}
\setlength{\tabcolsep}{4.5pt}
\renewcommand{\arraystretch}{1.12}
\begin{tabular}{@{}llll@{}}
\toprule
\textbf{Tool/Repository} & \textbf{\#1 (Most common)} & \textbf{\#2 (Second-most)} & \textbf{\#3 (Third-most)} \\
\midrule
sigstore/cosign         & CLI Tooling (30.3)              & Verification Workflow (17.5)        & Signing Workflow (17.5)             \\
sigstore/fulcio         & Policy/Configuration (22.2)     & Authentication/Authorization (15.4) & API (15.4)                          \\
sigstore/rekor          & API (39.1)                      & CLI Tooling (20.2)                  & Verification Workflow (8.5)$^{\dagger}$ \\
\midrule
keyfactor/signserver-ce & Web Client (25.0)               & Build/CI/Installation (17.5)        & Policy/Configuration (15.0)         \\
keyfactor/ejbca-ce      & Policy/Configuration (25.6)     & Web Client (24.0)                   & Build/CI/Installation (13.6)        \\
\midrule
notaryproject/notation  & CLI Tooling (36.2)              & Signing Workflow (15.6)             & Verification Workflow (12.4)        \\
openpubkey/openpubkey   & Authentication/Authorization (23.4) & API (14.5)                      & CLI Tooling (13.1)                  \\
hashicorp/vault         & Policy/Configuration (18.4)     & \makecell[l]{Key Management Core /\\ Secrets Backend (16.9)} & API (15.8)                      \\
\bottomrule
\end{tabular}

\bigskip
{\scriptsize $^{\dagger}$~For sigstore/rekor, \textit{Core} would rank \#3 (9.3\%);
Verification Workflow is shown as the next specific surface.
For all other tools, \textit{Core} ranked 4th or lower and did not affect the top-three listing.}
\end{table*}

\subsection{Results for RQ3}
\label{sec:appendix_results_rq3}

RQ3 asks how the reported usability problems change over time.
We include several additional figures and tables here, which are discussed in~\cref{sec:results_rq3}.

\begin{figure*}[t]
    \centering
    \includegraphics[width=0.99\linewidth]{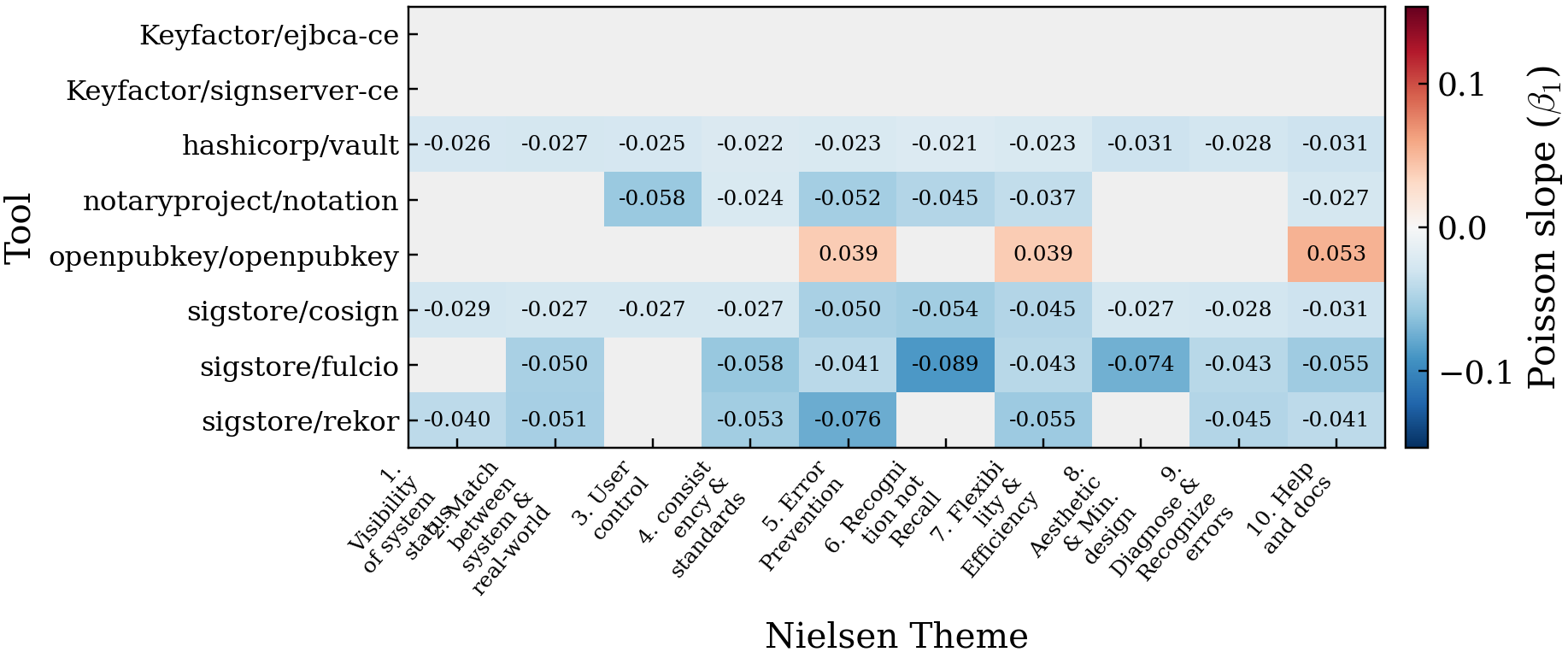}
    \caption{    
    Heatmap of Poisson time-trend slopes ($\beta_1$) by tool and Nielsen usability heuristic.
    Negative values indicate decreasing expected monthly issue counts over time; positive values indicate increasing counts.
    Only statistically significant cells ($p<0.05$) are annotated with the estimated $\beta_1$; non-significant cells are masked.
    }
    \label{fig:rq3-heatmap-nielsen}
\end{figure*}

\begin{figure}[t]
    \centering
    \includegraphics[width=0.95\linewidth]{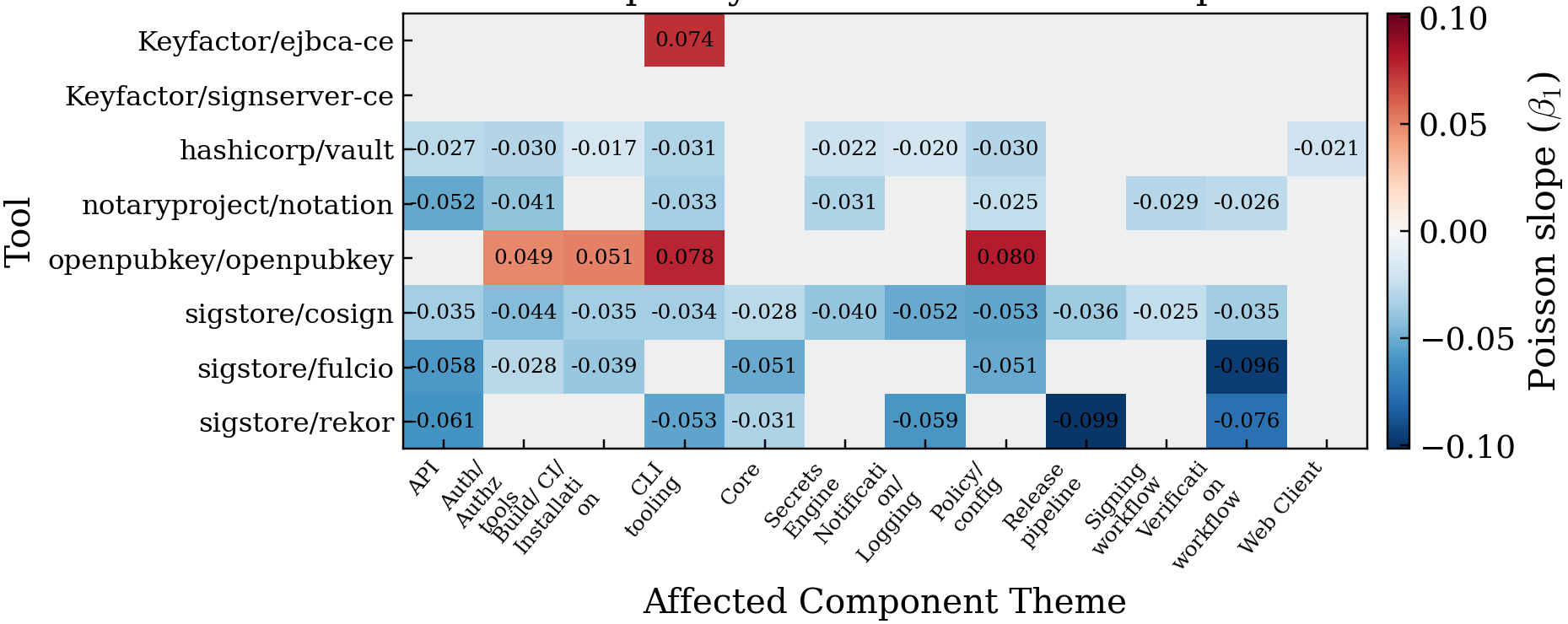}   
   \caption{Poisson time-trend slopes ($\beta_1$) for monthly usability-issue counts by tool and affected surface. Cells show statistically significant estimates ($p<0.05$); negative slopes indicate declining expected issue counts over time, positive slopes indicate increasing counts, and color intensity reflects slope magnitude.
    }
    \label{fig:heatmap-beta-associated_component_theme}
\end{figure}

\begin{itemize}[leftmargin=*, itemsep=3pt, topsep=2pt]
\item \cref{fig:rq3-heatmap-nielsen} shows Nielsen heuristic time-trend slopes by tool, indicating that declines in reported issues vary across heuristic categories and tools. This figure complements~\cref{fig:rq3-heatmap-l1-sig}.
\item \cref{tab:poisson-theme-family-by-tool} reports tool-level trends across the four theme families, including the Notation--Functional Reliability result discussed in~\cref{sec:results_rq3}.
\item \cref{tab:poisson-aggregate-l1-theme} reports aggregate Poisson time-trend estimates by inductive theme, showing that most themes decline over time but at different rates.
\item \cref{tab:poisson-aggregate-component} presents aggregate Poisson trends by affected surface, illustrating that some affected surfaces decline faster than others.
\item \cref{fig:heatmap-beta-associated_component_theme} visualizes tool-level Poisson slopes by affected surface, highlighting heterogeneous temporal trajectories across tools and surfaces.
\end{itemize}

\begin{table*}[t]
\centering
\caption{Repository-level Poisson time-trend estimates by theme family. Each cell reports $\beta_1$ with the classical Poisson $p$-value in parentheses; bold values indicate $p<0.05$. The 32 tests are interpreted as exploratory, with emphasis on direction, magnitude, and consistency across related models.}
\label{tab:poisson-theme-family-by-tool}
\scriptsize
\footnotesize
\setlength{\tabcolsep}{4pt}
\begin{tabular}{lrrrr}
\toprule
\textbf{Repository} & \textbf{Cognitive Friction} & \textbf{Functional Gap} & \textbf{Functional Reliability} & \textbf{Operational Friction} \\
\midrule
Keyfactor/ejbca-ce & $-0.0023$ (.745) & $0.0107$ (.471) & $0.0090$ (.414) & $0.0041$ (.574) \\
Keyfactor/signserver-ce & $0.0025$ (.888) & $0.0323$ (.337) & $0.0100$ (.807) & $0.0089$ (.652) \\
hashicorp/vault & $\mathbf{-0.0272}$ ($<$.001) & $\mathbf{-0.0245}$ ($<$.001) & $\mathbf{-0.0223}$ ($<$.001) & $\mathbf{-0.0237}$ ($<$.001) \\
notaryproject/notation & $\mathbf{-0.0162}$ (.029) & $\mathbf{-0.0358}$ ($<$.001) & $\mathbf{-0.0224}$ (.028) & $\mathbf{-0.0250}$ ($<$.001) \\
openpubkey/openpubkey & $\mathbf{0.0489}$ (.007) & $\mathbf{0.0347}$ ($<$.001) & $\mathbf{0.0431}$ (.005) & $\mathbf{0.0569}$ ($<$.001) \\
sigstore/cosign & $\mathbf{-0.0317}$ ($<$.001) & $\mathbf{-0.0500}$ ($<$.001) & $\mathbf{-0.0337}$ ($<$.001) & $\mathbf{-0.0313}$ ($<$.001) \\
sigstore/fulcio & $\mathbf{-0.0489}$ ($<$.001) & $\mathbf{-0.0473}$ ($<$.001) & $\mathbf{-0.0420}$ ($<$.001) & $\mathbf{-0.0413}$ ($<$.001) \\
sigstore/rekor & $\mathbf{-0.0326}$ ($<$.001) & $\mathbf{-0.0576}$ ($<$.001) & $\mathbf{-0.0575}$ ($<$.001) & $\mathbf{-0.0455}$ ($<$.001) \\
\bottomrule
\end{tabular}
\end{table*}

\begin{table}[t]
\centering
\caption{
Aggregate Poisson regression trends in expected monthly usability-issue counts by inductive theme.
Coefficients, rate ratios, and percentage changes are Poisson estimates; $p$-values use HC0 robust standard errors.
All trends remain negative and significant under negative-binomial sensitivity models.
Rows are sorted by $\beta_1$.
}
\label{tab:poisson-aggregate-l1-theme}
\footnotesize
\scriptsize
\setlength{\tabcolsep}{4.5pt}
\renewcommand{\arraystretch}{1.12}
\begin{tabular}{@{}lrrrrr@{}}
\toprule
\textbf{Theme} & \textbf{$\beta_1$} & \textbf{RR/Month} & \textbf{\%/Month} & \textbf{HC0 $p$} & \textbf{Total} \\
\midrule
Missing Feature/Enhancement Request & \textbf{-0.0326} & \textbf{0.968} & \textbf{-3.21} & \textbf{$1.47{\times}10^{-31}$} & \textbf{1,481} \\
Security Concerns & \textbf{-0.0326} & \textbf{0.968} & \textbf{-3.20} & \textbf{$1.96{\times}10^{-10}$} & \textbf{248} \\
Authentication Friction & \textbf{-0.0304} & \textbf{0.970} & \textbf{-2.99} & \textbf{$1.75{\times}10^{-4}$} & \textbf{199} \\
Tedious Workflows & \textbf{-0.0284} & \textbf{0.972} & \textbf{-2.80} & \textbf{$4.99{\times}10^{-10}$} & \textbf{401} \\
User Confusion/Unclear Documentation & \textbf{-0.0270} & \textbf{0.973} & \textbf{-2.66} & \textbf{$7.30{\times}10^{-14}$} & \textbf{982} \\
Integration Failure/Issues & \textbf{-0.0218} & \textbf{0.978} & \textbf{-2.16} & \textbf{$8.43{\times}10^{-11}$} & \textbf{659} \\
Unexpected Behavior & \textbf{-0.0218} & \textbf{0.978} & \textbf{-2.16} & \textbf{$8.28{\times}10^{-8}$} & \textbf{709} \\
Performance Issues & \textbf{-0.0218} & \textbf{0.978} & \textbf{-2.16} & \textbf{$0.0023$} & \textbf{97} \\
Notification/Logging Issues & \textbf{-0.0217} & \textbf{0.979} & \textbf{-2.14} & \textbf{$1.37{\times}10^{-6}$} & \textbf{512} \\
Build/CI/Installation/Distribution/Release Issues & \textbf{-0.0193} & \textbf{0.981} & \textbf{-1.91} & \textbf{$6.31{\times}10^{-5}$} & \textbf{272} \\
Configuration Friction & \textbf{-0.0193} & \textbf{0.981} & \textbf{-1.91} & \textbf{$5.60{\times}10^{-7}$} & \textbf{364} \\
\bottomrule
\end{tabular}
\end{table}

\begin{table}[t]
\centering
\caption{Aggregate Poisson regression slopes for expected monthly usability-issue counts by affected surface across all tools. Monthly counts are modeled in inclusive calendar-month bins over November 2021--November 2025. Coefficients, rate ratios, and percentage changes are estimated from the Poisson models; p-values are based on HC0 heteroskedasticity-robust standard errors. All 12 trends remain negative and statistically significant under negative-binomial sensitivity models. Rows are sorted by $\beta_1$ (steepest decline first).}
\label{tab:poisson-aggregate-component}
\footnotesize
\scriptsize
\setlength{\tabcolsep}{4.5pt}
\renewcommand{\arraystretch}{1.12}
\begin{tabular}{@{}lrrrrr@{}}
\toprule
\textbf{Affected Surface} & \textbf{$\beta_1$} & \textbf{RR/Month} & \textbf{\%/Month} & \textbf{HC0 $p$} & \textbf{Total} \\
\midrule
Verification Workflow & \textbf{-0.0338} & \textbf{0.967} & \textbf{-3.32} & \textbf{$4.20{\times}10^{-10}$} & \textbf{390} \\
CLI Tooling & \textbf{-0.0308} & \textbf{0.970} & \textbf{-3.03} & \textbf{$4.99{\times}10^{-18}$} & \textbf{1,110} \\
API & \textbf{-0.0306} & \textbf{0.970} & \textbf{-3.01} & \textbf{$6.06{\times}10^{-14}$} & \textbf{654} \\
Policy/Configuration & \textbf{-0.0289} & \textbf{0.971} & \textbf{-2.85} & \textbf{$1.33{\times}10^{-28}$} & \textbf{847} \\
Signing Workflow & \textbf{-0.0253} & \textbf{0.975} & \textbf{-2.50} & \textbf{$9.08{\times}10^{-7}$} & \textbf{407} \\
Authentication/Authorization & \textbf{-0.0246} & \textbf{0.976} & \textbf{-2.43} & \textbf{$2.59{\times}10^{-6}$} & \textbf{495} \\
Key Management Core / Secrets Backend & \textbf{-0.0241} & \textbf{0.976} & \textbf{-2.38} & \textbf{$4.16{\times}10^{-12}$} & \textbf{563} \\
Core$^{\dagger}$ & \textbf{-0.0227} & \textbf{0.978} & \textbf{-2.24} & \textbf{$2.86{\times}10^{-4}$} & \textbf{173} \\
Build/CI/Installation & \textbf{-0.0221} & \textbf{0.978} & \textbf{-2.19} & \textbf{$3.08{\times}10^{-7}$} & \textbf{340} \\
Notification/Logging & \textbf{-0.0221} & \textbf{0.978} & \textbf{-2.19} & \textbf{$1.91{\times}10^{-6}$} & \textbf{254} \\
Release Pipeline & \textbf{-0.0191} & \textbf{0.981} & \textbf{-1.89} & \textbf{$0.0019$} & \textbf{117} \\
Web Client & \textbf{-0.0161} & \textbf{0.984} & \textbf{-1.60} & \textbf{$2.18{\times}10^{-4}$} & \textbf{372} \\
\bottomrule
\end{tabular}
\bigskip\\
{\scriptsize $^{\dagger}$\textit{Core} represents issues where no specific sub-component boundary could be identified (solo-only assignments, $N=173$). 
}
\end{table}

\end{document}